\documentclass[%
reprint,
 amsmath,amssymb,
prb,
]{revtex4-2}
\usepackage{dcolumn}
\usepackage{bm}
\usepackage[pdftex]{graphicx}
\usepackage{amsmath}
\usepackage{amsfonts}
\usepackage{amssymb}
\usepackage{subfigure}
\usepackage{color}
\usepackage{float}
\usepackage[dvipsnames]{xcolor}
\usepackage{environ}
\NewEnviron{resize}{%
	\scalebox{0.9}{$\BODY$}
}
\usepackage{url}
\usepackage[linktocpage=true]{hyperref}
\usepackage{placeins}
\usepackage{soul}
\usepackage{float}
\usepackage{appendix}
\usepackage{indentfirst}

\begin{document}

\preprint{APS/123-QED}



\title{Symphony of Symmetry Selective Resonances in Fe-MgO-ZnO-MgO-Fe}


\author{Sabarna Chakraborti$^1$}
\author{Arti Kashyap$^2$}
\author{Abhishek Sharma$^{3}$}
\email{abhi@iitmandi.ac.in}
\affiliation{%
  1. Department Electrical Engineering, Indian Institute of Technology Ropar (Punjab, India)
}%
\affiliation{%
  2. School of Physical Sciences, Indian Institute of Technology Mandi (Himachal Pradesh, India)
}%
\affiliation{%
  3. School of Computing and Electrical Engineering, Indian Institute of Technology Mandi (Himachal Pradesh, India)
}%


\begin{abstract}
   \noindent We propose the perspective of symmetry-selective resonance of the $\Delta_1$ states in the Fe/MgO/ZnO/MgO/Fe heterostructures, offering a broad landscape to design magnetic tunnel junctions (MTJs) that yield a towering tunnel magnetoresistance (TMR) upto 3.5$\times10^4$\% with the resistance area (RA) product dipping down till a minimum of 0.05$\Omega$-$\mu m^2$, while maintaining a nearly perfect (99\%) spin polarization. Our predictions are based on the self-consistent coupling of the non-equilibrium Green’s function with the density functional theory. We also present the charge current, spin current and TMR of a prototype(p) Fe/MgO(3-layer(l))/ZnO(3l)/MgO(3l)/Fe-MTJ(abbreviated as p-$\Delta_1^R$-MTJ), which provides a pronounced advancement in the TMR(1.3$\times10^4$\%), RA product(0.45$\Omega$-$\mu m^2$) and spin polarization(99\%) over its Fe/MgO(6l)/Fe-based counterpart(TMR $\approx3.4\times10^3$\%, RA product$\approx22$$\Omega$-$\mu$$m^2$). Subsequently, we provide a holistic insight into the physics of symmetry-selective resonance by integrating the transmission eigenchannels, spectral density, and band structure of the Fe contacts associated with the p-$\Delta_1^R$-MTJ.
\end{abstract}

\maketitle

\section{\label{sec:level1}INTRODUCTION}
The story of magnetic tunnel junctions (MTJs) dates back to the discovery of tunnel magnetoresistance (TMR) in 1975\cite{JULLIERE1975225}, initiating a wave of research, that in due course, laid the theoretical foundations of Fe-MgO-Fe-based MTJs\cite{PhysRevB.63.054416}, followed by their successful fabrication in 2004\cite{parkin2004giant}.
\begin{figure*}[t]
    \centering
    \includegraphics[width= 16 cm]{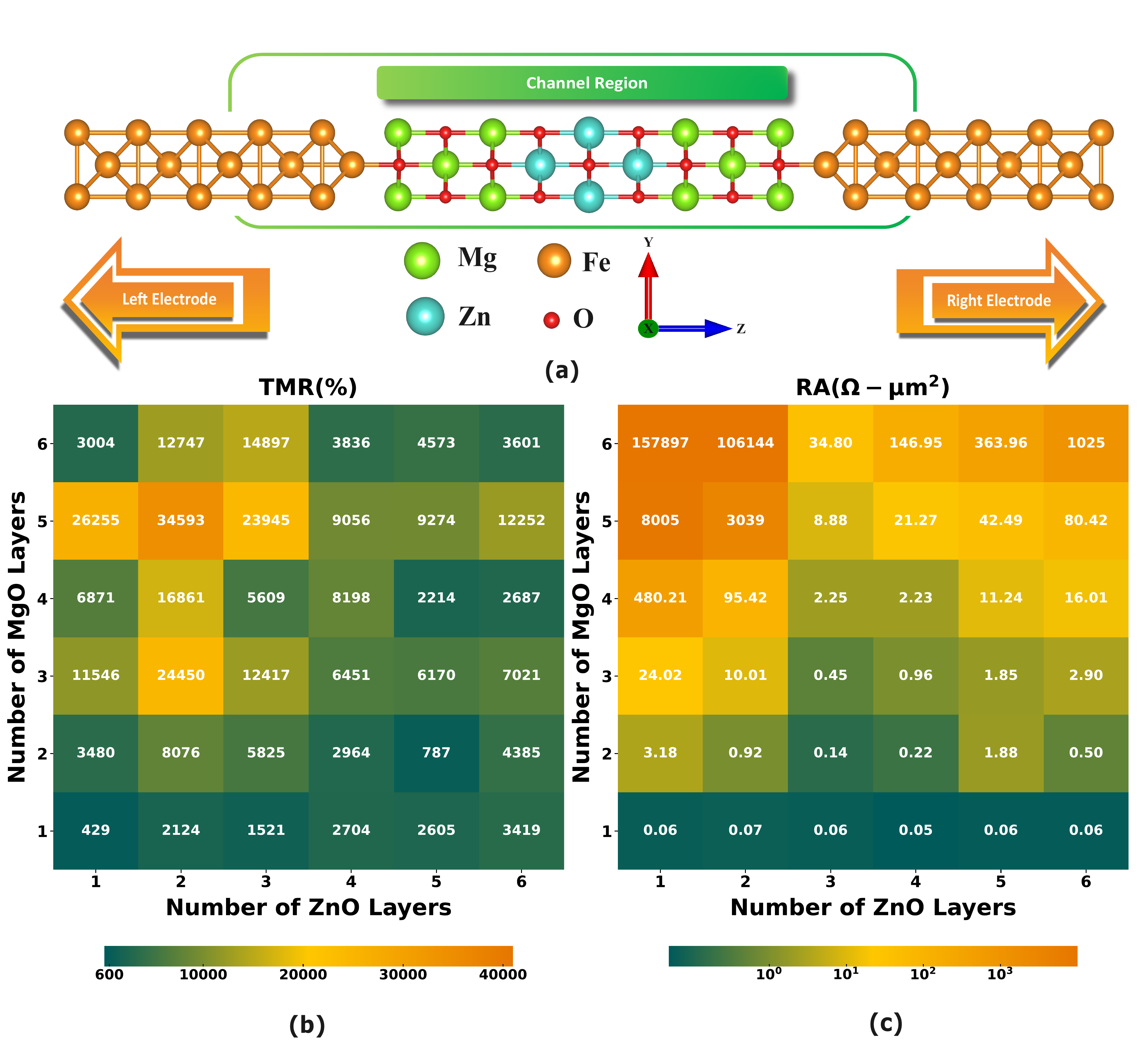} 
    \caption{(a)Schematic of two probe Fe (100)/ MgO (100)/ ZnO (100) /MgO (100) / Fe (100) MTJ. The device extends with infinite periodicity in both the $x$ and the $y$ direction with a lattice constant of 2.842 $\text{\AA}$ and the Fe electrodes extend upto $\pm\infty$ to in the $z$-direction. Landscape of the (b) TMR and the (c) RA product in the PC at zero bias with the number of MgO and ZnO layers. Here, an element with the n$th$ row and m$th$ column represents a $\Delta^R_1$-MTJ featuring an n-MgO/m-ZnO/n-MgO sandwich nestled between the Fe layers.}
q
    \label{fig:table}
\end{figure*}
A regular(r)-MTJ is made up of a free and fixed ferromagnet(FM) separated by a thin insulator(MgO)\cite{butler2001spin, waldron2006first}. Depending on the relative alignment of magnetization of the FMs, MTJs exhibit different resistances in the parallel ($R_{PC}$) and anti-parallel ($R_{APC}$) configurations. This behavior is governed by spin-dependent electron tunneling and is quantified by the TMR(\%)=$(R_{APC}-R_{PC})/R_{PC} \times 100$\%. The realization of purely electrical switching of the free FM through spin transfer torque (STT)\cite{berger1996emission, slonczewski1996current} positions the MTJs as a building block of energy-efficient circuits. The confluence of a high TMR, high STT (large spin polarization) and low resistance area product (RA) is imperative for MTJ-based spintronic devices. This synergy fosters swift and energy-efficient switching, with steadfast consistency, thereby making them top-tier contenders for a wide range of applications including sensing \cite{7124469, 9976889, 10.1093/nsr/nwx082}, neuromorphic computing\cite{Ielmini_2020, grollier2020neuromorphic, zhang2023spintronic, zhang2024domain}, probabilistic computing\cite{camsari2019pbits}, data storage\cite{ 10.1063/1.5129902, cardoso2014magnetic, liu2024domain}, non-volatile magnetic logics\cite{9235579} and ambient energy harvesters\cite{sharma2024nanoscale} to name a few. The simple design, high density, affordability and reliability of STT-magnetoresistive random access memories(MRAMs) also make them suitable candidates for embedded flash memories\cite{nature2020sttram,dieny2020opportunities,hu2022dsmtj,han2024gpu, han2024cpu, GF2023}. While the roadmap to achieve a density of 100 $Gbit/cm^2$ STT-MRAM is outlined with ultra-scaled MTJs\cite{10368042}, the aggressive scaling is found to increase the RA product and reduce the TMR\cite{watanabe2018shape}. 

Besides, the low latency, non-volatile spin-orbit torque(SOT)-MRAMs carry the potential of replacing static (S)-RAMs in the L1 cache\cite{imec2022sotmram, nguyen2024sotmram}. Similarly, the skyrmionic memories, residing in the ambit of research, hold significant promise for further energy-efficient and denser data storage\cite{shao2024magnetic,rajib2024skyrmion,8667872}. In SOT-MRAMs and skyrmionic memories, MTJs serve as the key component by enabling the electrical readout of magnetization through the TMR effect. Hence, high-TMR and low RA MTJs offer a pivotal role to transform the semiconductor memory landscape, delivering superior speed, energy efficiency, and improved scalability for next-generation computing.
\begin{figure*}[t]
    \centering
    \includegraphics[width= 17 cm]{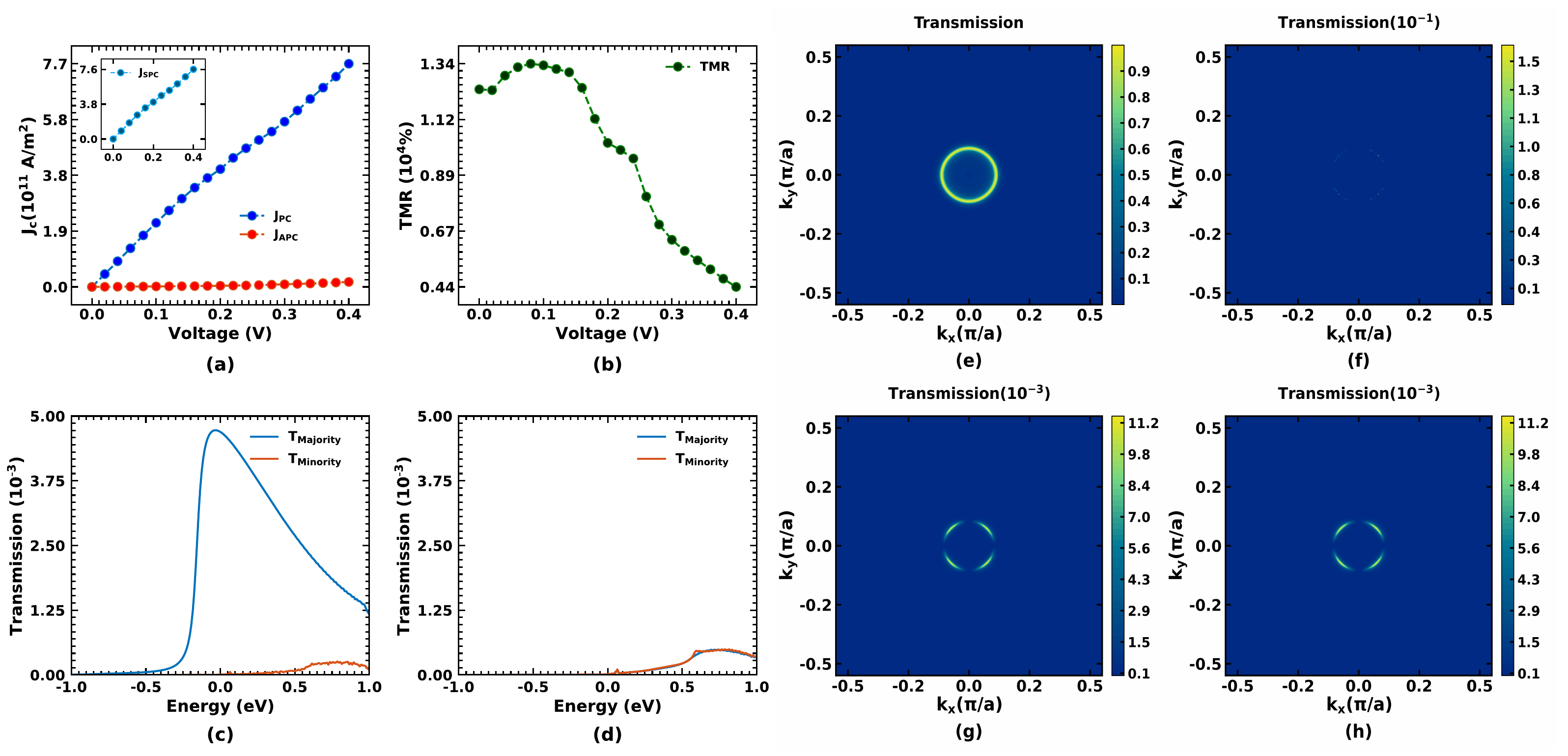} 
    \caption{Variation of the (a) Current density($J_C$)[Inset: J$_{SPC}$ in PC] and (b) TMR\% of the p-$\Delta_1^R$-MTJ with the applied voltage. k-averaged transmission of the p-$\Delta^R_1$-MTJ in the (c) PC and (d) APC at zero bias with Fermi energy $E_f=0$ eV. k-resoved transmissions of the p-$\Delta^R_1$-MTJ for the (e) majority and the (f) minority electrons in the PC, the (g) majority and the (h) minority electrons in the APC at Fermi energy with no applied bias.}
    \label{fig:TM_IV}
\end{figure*}
Technologically relevant advancements in spintronics towered on MTJs, trace their origin back to the seminal work on the Fe/MgO/Fe sandwich that predicted a TMR of over 1000\%\cite{PhysRevB.63.054416}. Owing to the symmetry dependent tunneling, Fe-MgO-Fe experimentally yielded a TMR of 220\% under regular conditions while climbing to over 500\% in the chill of lower temperatures \cite{yuasa2004giant, parkin2004giant}. The MgO-based MTJs still fell short, with recent developments managing to push CoFeB/MgO/CoFeB MTJs to a TMR of 631\%\cite{scheike2023room}. The theoretical limit of symmetry-dependent tunneling in MgO-based MTJs has prompted the investigation of alternative electrodes\cite{yang2018ultrahigh} and channels using 2D van der Waals (vdW) \cite{balci2018high, zhu2023large, halder2024half} and 3D-heterostructure-based MTJs \cite{nawa2021enhanced}. However, the pursuit of simultaneously achieving high TMR, low RA, and high STT remains stalled. Challenges such as intricate fabrication processes, weak CMOS compatibility, and reliability continue to hinder the development of MTJs based on novel materials.

Considering the recent development of the STT-MRAM technology at the 14 nm node\cite{lee2022world}, we present a design of heterostructure-based MTJs to enhance the performance metrics simultaneously, while retaining the fabrication maturity developed over decades for the FM–MgO interfaces. In this work, we explore the symphony of resonance in the broad landscape of the Fe/MgO/ZnO/MgO/Fe-based MTJs($\Delta^R_1$-MTJs), through the $\Delta_1$ symmetry states via first-principle calculations. The choice of ZnO stems from its favorable conformation in the rocksalt($rs$) state with a minimal lattice mismatch of 1.90\%\cite{10.1063/1.2188382} with $rs$-MgO. The nanocrystals of $rs$-ZnO can grow up to a thickness of 2 nm with a surprisingly low surface energy of 0.63 J/m$^2$\cite{koster2015stabilization}. The fabrication of $rs$-ZnO(001) on top of $rs$-MgO(001) at ambient conditions\cite{lu2016growth, eijt2007formation} further adds to the feasibility of MgO/ZnO heterostructure. Moreover, Y. Uehara et. al. reported the fabrication of a Fe(100)/($rs$-MgO(100)/$rs$-ZnO(100))$_n$/Fe-MTJ with low RA and TMR in comparison to regular MTJ (Fe(100)/$rs$-MgO(100)/Fe)\cite{uehara2005mtj}. Additionally, H. Saito et. al. has also demonstrated the successful fabrication of a spin-dependent diode comprised of Fe/$rs$-MgO(100)/$rs$-ZnO(100)/Fe for low RA with asymmetric I-V characteristics and a moderate TMR\cite{PhysRevApplied.11.064032}. Although these works demonstrate experimental realization of $rs$-MgO(100)/$rs$-ZnO(100) on Fe, but, they have missed the broad design landscape of the Fe/rs-MgO/rs-ZnO/rs-MgO/Fe-based MTJs, where both Fe layers are in direct contact with MgO layers. This seemingly trivial but crucial design modification gives rise to the intriguing physics of symmetry-selective resonant tunneling capable of delivering high-performance $\Delta^R_1$-MTJs.
We chronicle this study by presenting a brief overview of the results, offering a broad spectrum of TMR reaching up to 10 times and the RA product being reduced to approximately 400 times that of a r-6 layers(l)-MTJ and subsequently elaborate a case study of the Fe/MgO(3l)/ZnO(3l)/MgO(3l)/Fe-based $\Delta^R_1$-MTJ. After that, we lay down a complete roadmap for curating the $\Delta^R_1$-MTJs with a wide range of performance metrics, such as the TMR and the RA product, and the trade-offs involved among them.

\section{Methodology}
We utilize self-consistently coupled the density functional theory (DFT) with the non-equilibrium Green's function (NEGF) to perform the spin-polarized transport calculations across the $\Delta^R_1$-MTJ shown in Fig.~\ref{fig:table}(a). We employ the Perdew-Burke-Ernzerhof(PBE) generalized gradient approximation(GGA) for the exchange-correlation functional and the standard norm-conserving pseudopotentials\cite{Soler2002} while using the double zeta LCAO basis set for s, p and d orbitals using the Siesta framework\cite{VANSETTEN201839, zerothi_sisl}. We also benchmarked the conventional Fe/MgO/Fe-based MTJ (see \hyperref[appendix A]{Appendix A}) to demonstrate that transmission eigenchannels\cite{Madseigen} offer a more generalized approach to rationalize symmetry-dependent tunneling in MTJs, compared to the complex band structures\cite{butler2001spin, nawa2021enhanced}.

\begin{figure*}[t]
    \centering
    \includegraphics[width= 17 cm]{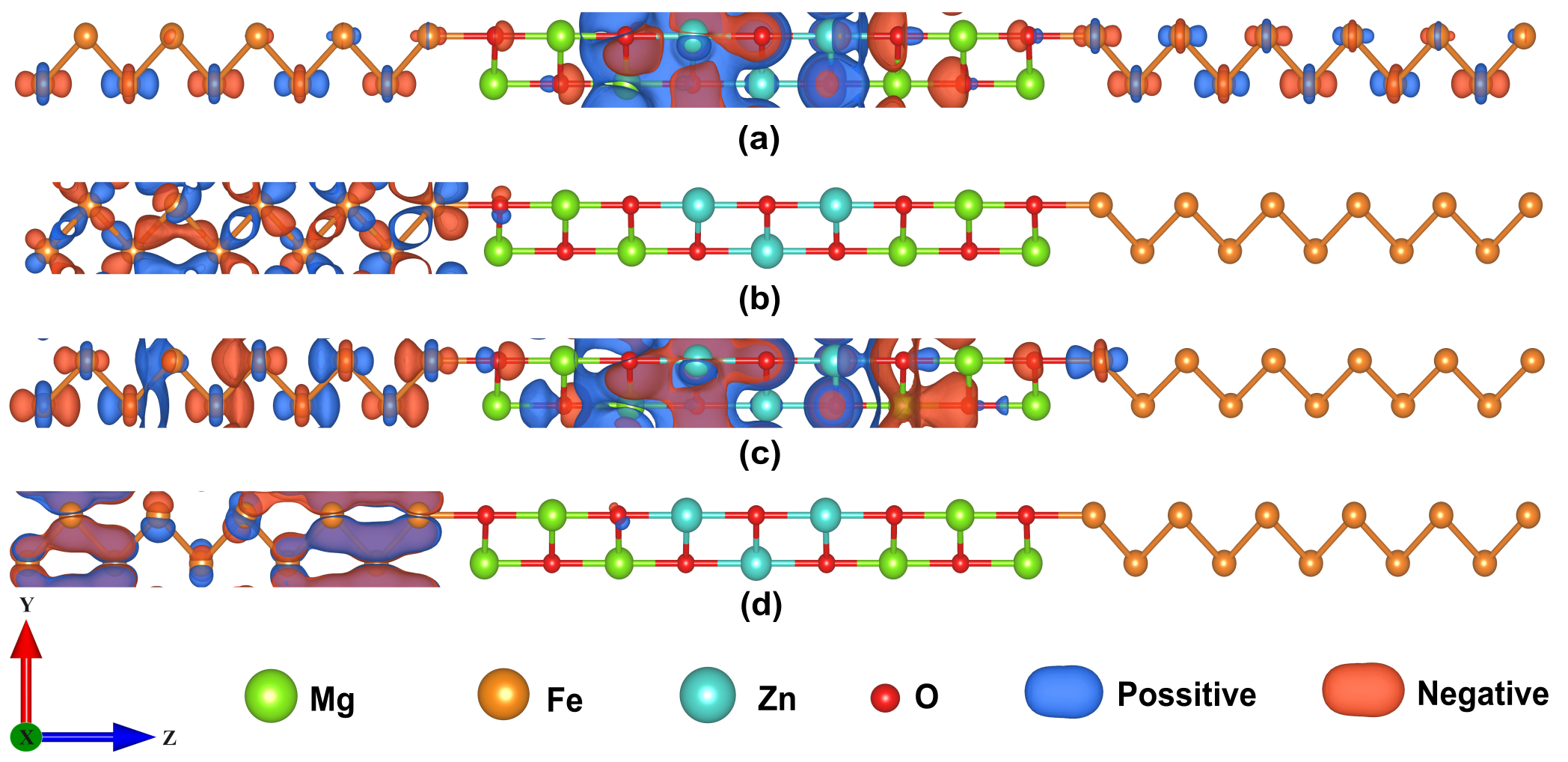} 
    \caption{c-eigenchannels of the p-$\Delta^R_1$-MTJ, emanating from the left electrode for the (a) majority and the (b) minority electrons in the PC with $\lvert k_x \lvert= 0$, $\lvert k_y \lvert= 0.11\pi / a$, followed by (c) the majority and (d) the minority electrons in the APC with $\lvert k_x \lvert = 0.08\pi / a$, $\lvert k_y \lvert= 0.06\pi/a$ at the Fermi energy with zero applied bias. The blue and red isosurfaces represent eigen wavefunctions with isovalues 0.1 and -0.1, respectively.}
    \label{fig:C_333}
\end{figure*}

The recent study on r-MTJs showed that GGA-PBE functionals are adequate for accurately predicting the transmission of the majority electrons\cite{PhysRevB.110.224425}. The GW calculation does not alter the transmission of the Fe/MgO/Fe-majority states\cite{PhysRevB.110.224425, PhysRevB.85.174433}, possibly due to the population of the de-localized states (s-orbitals) in the MgO layers injected by majority electrons of the Fe electrode(see \hyperref[appendix D]{Appendix D}. The GW calculation of the r-MTJ shifts the interface resonance states of the minority electrons only by 0.12 eV, leading to an increase in TMR near the zero bias\cite{PhysRevB.110.224425}. Since the $\Delta_1^R$-MTJs rely on the symmetry-selective resonance of majority-spin electrons ($\Delta_1$ states, see Fig.~\ref{fig:C_333}(a)), the predicted TMRs are at best underestimated. Likewise, the predicted RA (in PC) values, being dominated by symmetry-selective resonant tunneling of majority electrons, are unlikely to be significantly affected. The presence of low spin orbit coupling(SOC) at both the Fe-MgO boundary and in ZnO ensures prolonged spin coherence, which leads to longer spin relaxation lengths and effectively neutralizes any impact of SOC on the $\Delta_1^R$-MTJ\cite{PhysRevB.105.195202, PhysRevX.7.031018}. Hence, the GW correction and SOC are neglected to the first order to manage the computational cost. A thorough evaluation of these aspects may be pursued in future studies.

We start with relaxing the geometry of the $\Delta_1$-MTJs using density functional theory (DFT), implemented in the Vienna Ab initio simulation package (VASP) [58,59] with a k-grid of 20$\times$20$\times$1 and kinetic energy cutoff of 600 eV. Further, the geometry of the $\Delta^R_1$-MTJs (such as Fig.~\ref{fig:table}) is relaxed by sampling the transverse k-grid with 20$\times$20($k_x$, $k_y$) k-points with a cutoff energy of 600 Ry for the double $\zeta$-polarized basis set in Siesta, and a decent agreement is observed. The electrodes were fixed at crystalline positions during the relaxation. The density matrix was relaxed to an accuracy of 10$^{-3}$ till the Hellman-Feynman forces dropped below 0.04 eV/$\AA$. During the geometry optimization, the lattice constant of the device in the transverse direction is held to that of the bulk of BCC-Fe ($\approx$2.847 $\AA$)\cite{huang2021thickness}. The Hamiltonian matrices for the Fe electrodes are generated employing a Monkhorst grid of 15$\times$15$\times$100. The convergence of the non-equilibrium density matrix on the complex energy contour is achieved by sampling the Brillouin zone (BZ) with a \( 15 \times 15 \) (\( k_x, k_y \)) k-point grid for DFT + NEGF calculations, ensuring self-consistency until the residual charge is reduced below \( 10^{-3} \) \cite{PhysRevB.65.165401}. However, a far denser k-grid of 1200$\times$1200($k_x$, $k_y$) is required to capture the transmission spectra in the first BZ. The channel region is designed with Fe(11l) / MgO(3l) / ZnO(3l) / MgO(3l) / Fe(10l)) to ensure smooth boundary conditions with Fe electrodes. The mathematical framework used for the NEGF+DFT calculation is described in  \hyperref[appendix E]{Appendix E}. 

\section{Results}

We commence by presenting the broad design landscape of Fe/MgO(ml)/ZnO(nl)/MgO(ml)/Fe $\Delta^R_1$-MTJs where n,m $\in$[1,6] in Fig.~\ref{fig:table}.
Fe is in the BCC phase, and both MgO and ZnO are in the rocksalt FCC phase. The $\Delta^R_1$-MTJs offer a wide range of zero bias TMR($788-34593\%$) and RA ($0.05-3038$ $\Omega\text{-}\mu m^2$)product with high spin polarization ($77.7\%-99.7\%$) in the PC. 
The wide design manifold featuring a superior performance triad of high TMR, low RA and high spin polarizations (leading to large STT) in conjunction with higher stability(see \hyperref[appendix C]{Appendix C}) is indicative of rendering the $\Delta^R_1$-MTJs, a superior alternative for a large number of applications over their regular counterparts \cite{imec2022sotmram,nguyen2024sotmram,Ielmini_2020, grollier2020neuromorphic, zhang2023spintronic, zhang2024domain}. 
We investigate a prototype(p)-$\Delta_1^R$-MTJ comprised of Fe(11l)/MgO(3l)/ZnO(3l)/MgO(3l)/Fe(10l) via transmission eigenchannels and explore the symmetry selective resonance in the $\Delta_1^R$-MTJs. The choice of the prototype is attributed to its optimized performance matrices in the form of the TMR(429\%-1.34$\times 10^4 \%$), the RA product(0.45--15.8$\times10^4$ $\Omega\text{-}\mu m^2$) and the high spin polarisation(99.7\%) in conjunction with the conformity to MgO-layer thicknesses that are experimentally feasible\cite{deac2008bias, patel2020mgo}. In this case, we achieve a maximum TMR that supersedes the r-MTJ by a factor of four with the RA product decimated by 50 times (see  \hyperref[appendix A]{Appendix A}). In the p-$\Delta^R_1$-MTJ, the distance between the  Fe-O atoms is found to be 2.198 $\AA$, 0.08$\AA$, shorter than the r-MTJ (2.228 $\AA$ see  \hyperref[appendix A]{Appendix A}). The $free\ energy$\cite{wentzcovitch1992energy} of the structure is found to be  $-21.86\times 10^3 eV$ lower compared to a regular 6-layer-based MTJ(see Appendix A\ref{appendix A}), reflecting higher thermodynamic stability possibly due to a stronger bond between Fe-O atoms in the $\Delta^R_1$-MTJ(see  \hyperref[appendix C]{Appendix C} for details).
In Fig.~\ref{fig:TM_IV}(a) we show the variation of the charge current density $J_C$, of the p-$\Delta^R_1$-MTJ for both the PC and the APC, in conjunction with the spin current density $J_{SPC}=J_{\uparrow}-J_{\downarrow}$ with the applied voltage.

\begin{figure*}[t]
    \centering
    \includegraphics[width= 16 cm]{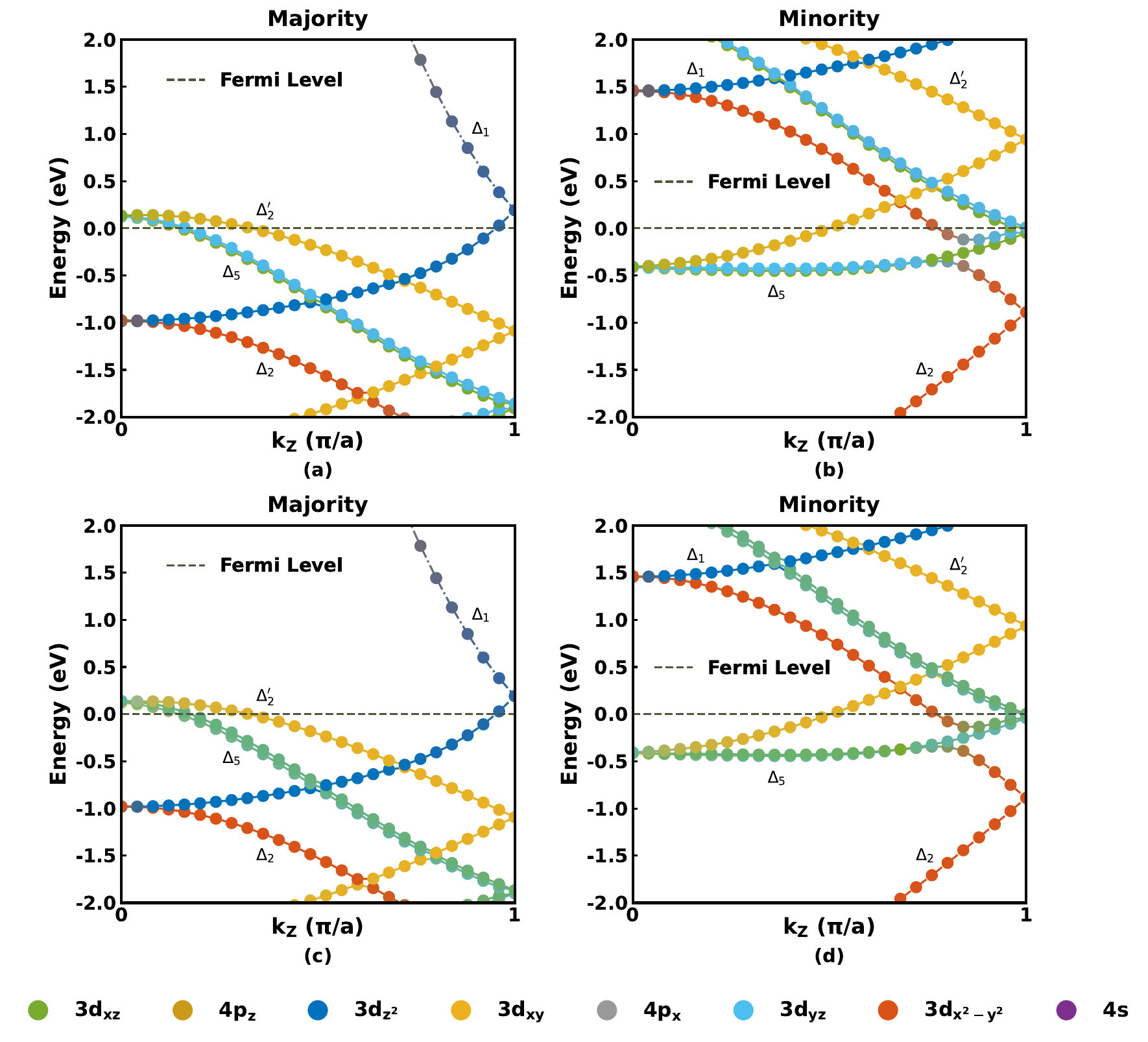} 
    \caption{The band structure of Fe electrodes are depicted for: (a) the majority and (b) the minority electrons at $\lvert k_x \rvert = 0$, $\lvert k_y \rvert = 0.11\pi / a$; and (c)the majority and (d) the minority electrons at $\lvert k_x \rvert = 0.08\pi / a$, $\lvert k_y \rvert = 0.06\pi/a$, as $\lvert k_z \rvert$ is varied from $0$ to $\pi /a$. The colour coding of the bands corresponds to the orbital composition, as illustrated in the colour bar. }  
    \label{fig:BD_333}
\end{figure*}

\begin{figure}[h]
    \centering
    \includegraphics[width= 8.5 cm]{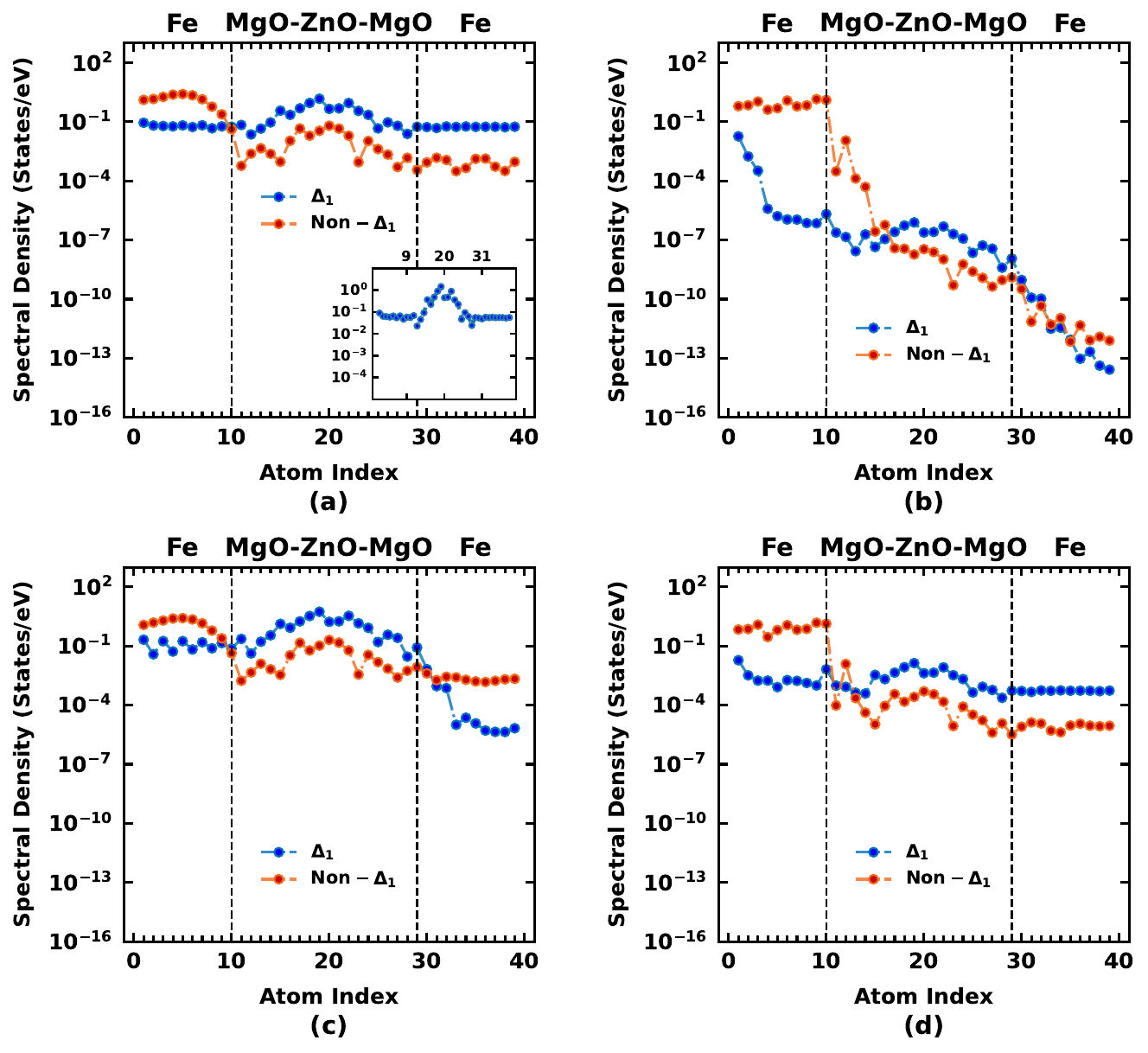} 
    \caption{Spectral density of the p-$\Delta^R_1$-MTJ, originating from the left electrode for (a) the majority and the (b) the minority electrons in the PC at $\lvert k_x \rvert = 0$, $\lvert k_y \rvert = 0.11 \pi/a$ followed by (c) the majority and (d) the minority electrons in the APC with $\lvert k_x \rvert = 0.08\pi/a$, $\lvert k_y \rvert = 0.06\pi/a$ at Fermi energy.}
    \label{fig:ADOS_333}
\end{figure}

The momentous boost in the transmission of the majority carriers in the PC gives rise to a towering TMR(\%) with a maximum value of 1.34$\times$10$^4\%$ along with an unprecedented reduction in the RA product(see Fig.~\ref{fig:TM_IV}(a) and (b)). Unlike a regular MgO-based MTJ, the TMR(\%) does not fall quite so rapidly\cite{waldron2006first} but shows a mild increase at first. In Fig.~\ref{fig:TM_IV}(c) and (d), we present the k-avg transmission of the p-$\Delta_1^R$-MTJ in the PC and the APC, respectively.
Probing into k-space of the $\Delta^R_1$-MTJ unveils a circular arc shown in Fig.~\ref{fig:TM_IV}(e), rendering a perfect transmission (termed as resonant transmission) of 1 with a radius, nearly in the order of 0.11$\pi/a$, where $a$ is the dimension of the unit cell associated to the contacts in the transverse direction. This, in turn, increases the overall k-averaged transmission of the majority electrons in the PC.
However, the transverse momentum space of the minority electrons in the PC (see Fig.~\ref{fig:TM_IV}(f)) and both the majority (see Fig.~\ref{fig:TM_IV}(g)) and the minority electrons (see Fig.~\ref{fig:TM_IV}(h)) in the APC, reverberate with off-resonance(or tunneling). The transmission in the k-space is deliberately given from -0.5$\pi/a$ to 0.5$\pi/a$, to present a zoomed landscape of the transport features, as the rest of the BZ did not contain any notable signature.
To explore the underlying reason for such behaviour, we delve into the orbital resolved band structure of the Fe electrode/contact (see Fig.~\ref{fig:BD_333}) and corroborate it with the conducting eigenchannels\cite{Madseigen} of the $p-\Delta^R_1$-MTJ. To closely follow the developments in the k-space, we commence our investigation with the band structure of the Fe electrode along the transport direction at $k_x=0$, $k_y=0.11\pi/a$ point, which lies on the bright ring of the transmission spectra (see Fig.~\ref{fig:TM_IV}(e)).
The band structure of the majority electrons in Fe primarily comprises $\Delta_1$(s, p$_z$ and d${_{Z^2}}$ orbitals), $\Delta_2'$( $d_{xy}$ orbital) and $\Delta_5$($d_{xz}$, $d_{yz}$ orbitals) states near the Fermi level(see Fig.~\ref{fig:BD_333}(a)). 
The majority electron states of the Fe contact at the Fermi level (with $k_x$=0 and $k_y= 0.11\pi/a$) with $\Delta_1$ symmetry are found to propagate through the Fe/MgO/ZnO/MgO/Fe heterostructure without any attenuation as demonstrated by the conducting (c-) eigenchannel of the p-$\Delta^R_1$-MTJ in Fig.~\ref{fig:C_333}(a)(see  \hyperref[appendix B]{Appendix B} for the non(n)-c eigenchannel). The $\Delta_1$ symmetry states seamlessly transits (termed as resonance of $\Delta_1$ states) through the heterostructure via coupling to the $p_z$ and the $s$ orbital of the O and Mg/Zn atoms, respectively. The presence of high spectral density(SD) of the $\Delta_1$ symmetry states with $k_x$=0 and $k_y= 0.11\pi/a$ at the Fermi level in the p-$\Delta_1^R$-MTJ, as shown in Fig.~\ref{fig:ADOS_333}(a), further corroborates its participation in the transport. 

The band structure of the minority electrons in Fe primarily comprises $\Delta_2$($d_{x^2-y^2}\ \text{orbital}$), $\Delta_2'$, and $\Delta_5$ states near the Fermi level(see Fig.~\ref{fig:BD_333}(b)). The c-eigenchannel for the minority electrons in the PC presented in Fig.~\ref{fig:C_333}(b), shows the coupling of the $d_{yz}$ orbital of the left contact to the $p_y$ orbital of the O atom, indicating the propagation of the $\Delta_5$ state of the left electrode(see Fig.~\ref{fig:BD_333}(b)) through the MgO. Nevertheless, the rapid decay of SD of the non-$\Delta_1$ symmetry states ($\Delta_2,\Delta^{'}_2, \text{and}\ \Delta_5$) as depicted in Fig.~\ref{fig:ADOS_333}(b) and the fading c-eigenchannel across the channel presented in Fig.~\ref{fig:C_333}(b), is indicative of regular tunneling (termed as off-resonant). 
The propagation of the majority c-eigenchannel in the APC, however, is more interesting. To begin our analysis, we plot the bandstructure of the Fe contact (Fig.~\ref{fig:BD_333}(c)) along the transport direction at $k_x=0.08 \pi/a$, $k_y=0.06 \pi/a$, a point of bright transmission signature in the first BZ(see Fig.~\ref{fig:TM_IV}(g)  and (h)). Here, we see the $\Delta_1$ states, dominated by the $d_{Z^2}$ orbitals to comfortably propagate via the channel region(see Fig.~\ref{fig:C_333}(c)). But the absence of the $\Delta_1$ band in the right electrode (Fig.~\ref{fig:BD_333}(d): majority spin of the left contact are the minority spin of the right contact in the APC and vice-versa) forces the electrons to get reflected. This phenomenon is strongly substantiated by the SD of the $\Delta_1$ state shown in Fig.~\ref{fig:ADOS_333}(c), which, although sustained at a high magnitude till the terminus MgO layer, falls off sharply as we move further in the right electrode.
The c-eigenchannel of the minority electrons in the APC shown in Fig.~\ref{fig:C_333}(d), primarily demonstrates the participation of the $d_{xz}$ orbitals near the Fe/MgO interface. However, the SD plotted in Fig.~\ref{fig:ADOS_333}(d) reveals the presence of the $\Delta_1$ state in the Fe/MgO/ZnO/MgO/Fe. This might look paradoxical at first glance, as the left electrode does not seem to have any $\Delta_1$-symmetry state at Fermi energy(see Fig.~\ref{fig:BD_333}(d)) for the minority electrons. Nonetheless, closely examining the minority spin fatbands of left contact indicates a minor presence of the $d_{Z^2}$ orbitals (see Fig.~\ref{fig:BD_333}(d): it barely alters the colour composition), and the presence of the $\Delta_1$ electrons in the right electrode (see Fig.~\ref{fig:BD_333}(c)) offer a leeway for the electrons with the respective symmetry.

\begin{figure}[h]
    \centering
    \includegraphics[width= 8 cm]{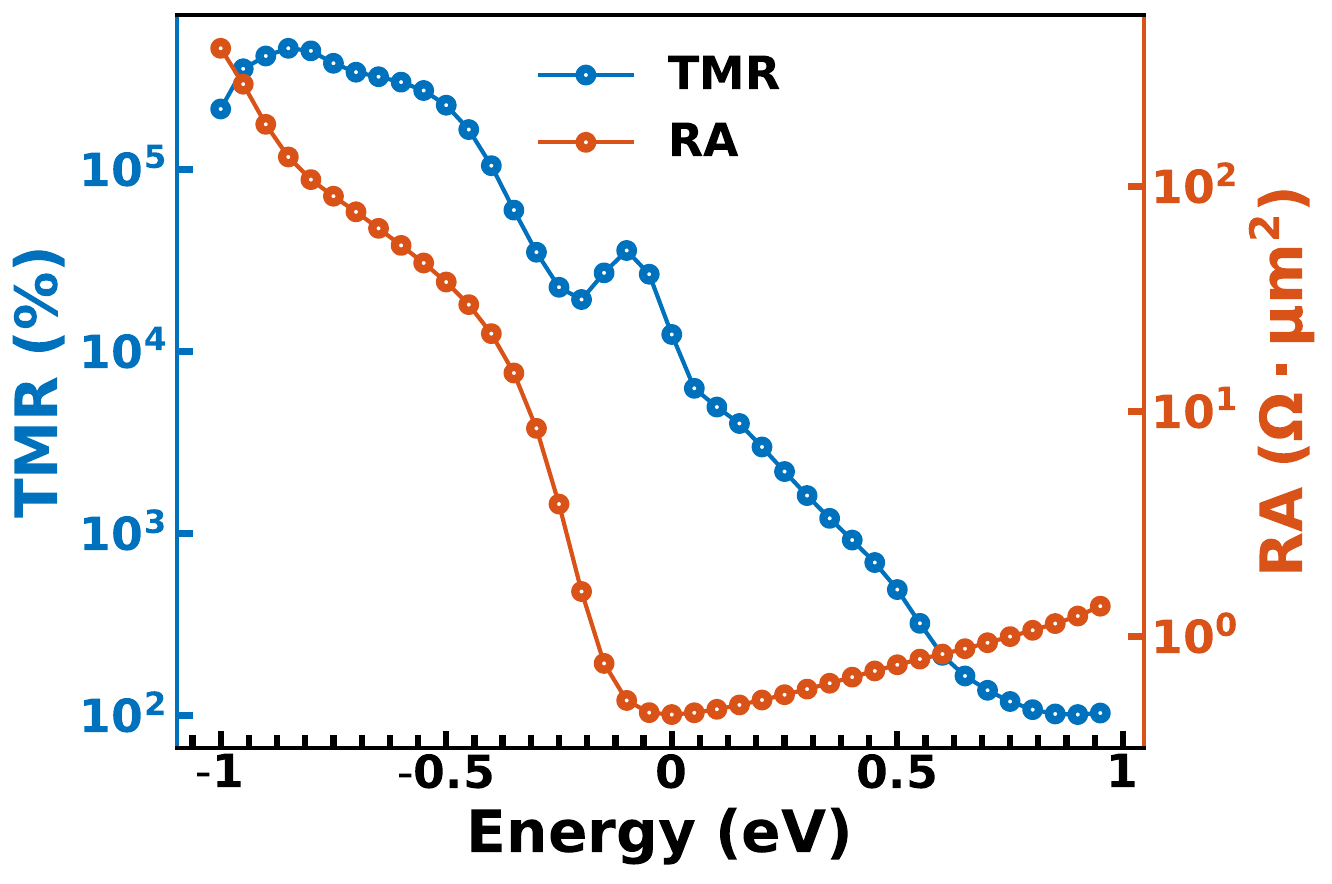} 
    \caption{Dependence of the TMR and the RA product in the PC of the p-$\Delta_1$-MTJ at zero bias with energy. Here, the energy axis along the $x$-direction is given with reference to the Fermi energy of the contacts.}
    \label{fig:TMR_Energy}
\end{figure}

To get an overview of the stability of the TMR\% and the RA product of the  p-$\Delta_1$-MTJ, we present their dependence on energy as shown in Fig.~\ref{fig:TMR_Energy}. It is found that both the TMR\% and RA product exhibit a notable robustness across a broad energy window around the Fermi level ($E_f=0$eV). The TMR\% shows a smooth trend without any abrupt fluctuations or resonance peaks, indicating that the spin-dependent tunnelling is governed by bulk electronic states rather than interface-localized states, which are typically sensitive to energy. A broad and smooth TMR and RA imply that even under thermal broadening or applied biases, the spin polarisation of the tunnelling current will remain largely unaffected, ensuring the stability with varying temperature and voltages, thereby reinforcing the suitability of the p-$\Delta_1$-MTJ for next-generation spintronic applications.

Finally, we elucidate the design landscape of the $\Delta^R_1$-MTJs presented in Fig.\ref{fig:table}(b) and (c). The $\Delta^R_1$-MTJs exhibit a general trend of oscillating TMR (\%) with increasing MgO/ZnO layers similar to MgO-based regular MTJs\cite{matsumoto2007oscillation,scheike2023631}.
At the same time, the RA product of the $\Delta^R_1$-MTJs in the PC surges monotonically as the thickness of the MgO increases, whereas increasing the number of ZnO layers first brings a sharp dip in RA and displays an oscillatory character afterwards. 
The oscillatory trend of the RA product and the TMR can be attributed to: (1) faster decay of the non-$\Delta_1$ states with an increase in MgO layers, and (2) emergence of perfect transmission channels for $\Delta_1$ states at different energies with varying ZnO layers. The interplay between MgO and ZnO layers creates a vast difference between the currents in the PC and the APC, providing a broad design space for $\Delta^R_1$-MTJs, as illustrated in Fig.~\ref{fig:table}. Accordingly, we hold the view that this development would enable the engineers to adopt case-specific designs, offering MTJs with a diverse range of TMR and RA tailored to specific requirements.
\section{Conclusion}
Since the discovery of FM/MgO/FM MTJs\cite{butler2001spin,waldron2006first},  two decades have passed in the search for materials that simultaneously enhance the TMR, lower the RA product, and offer a high spin polarization (for energy-efficient spin-transfer torque switching) in MTJs. Keeping aside the fabrication efforts, even the theoretical endeavours have struggled to yield the desired outcomes.
In this work, we present the phenomenon of symmetry selective resonance of the $\Delta_1$ states via first principle calculations in the Fe/MgO/ZnO/MgO/Fe heterostructures while working within the confines of today’s fabrication techniques\cite{patel2020mgo}. Modulating the thickness of the MgO and the ZnO layers gave a broad spectrum of TMR($788$-$34593\%$) and RA($0.05$-$3038$ $\Omega$-$\mu m^2$), which holds the potential to far exceed the bounds of an r-MTJ. Along with a large boost in the TMR, the unprecedented reduction in the RA products of the $\Delta_1^R$-MTJs in the PC indicates a sizable increase in the spin current owing to the dominance of the T$_{\text{Majority}}$. It is thus anticipated that advances in the fabrication of the $\Delta_1$-MTJs will pave the avenue for faster and energy-efficient spintronics in the coming days.  \\ \\
\section*{Authors Contributions:}A.S conceived the idea, conceptualized and guided the study. S.C performed the DFT + NEGF calculations and wrote the first draft of manuscript. A.S and S.C formulated the perspective of symmetry-selective resonant tunneling via eigen-channel analysis. A.K provided valuable DFT insight of the materials. All authors discussed the results and the manuscript.\\
\section*{Acknowledgement} The author Abhishek Sharma acknowledges the support by the Science and Engineering Research Board (SERB), Government of India, Grant No. SRG/2023/001327. A.S also acknowledges the valuable discussion about Siesta framework with Dr. Nick Papior, Dr. Vivekanand Shukla, Dr. Konrad Wilczynski and Dr. Naina Devraj. The authors also acknowledges Param Himalaya superconducting facility at IIT Mandi under National supercomputing-mission for valuable resource support. A.S acknowledges the valuable technical support provided by Sowmya Shree N, Mr. Samir Shaik and Mr. Anil Kumar on Param Himalaya supercomputing facility.\\
\section*{Data Availability:} The data that support the findings of this study are  \href{https://drive.google.com/drive/folders/1EB1wJahK4reGhTEDrzHPb64sswAk4BFY?usp=sharing}{available} upon reasonable request to the corresponding author.

\section*{Appendix A: Exploring the 6-layer-$\text{MgO}$-based MTJ}
\label{appendix A}
We venture into the study of MTJs by first benchmarking our results with the Fe-MgO-Fe sandwich demonstrated via first principle calculations\cite{butler2001spin, waldron2006first} with the existing literature. 
\begin{figure}[h]
    \centering
    \includegraphics[width= 8.3 cm]{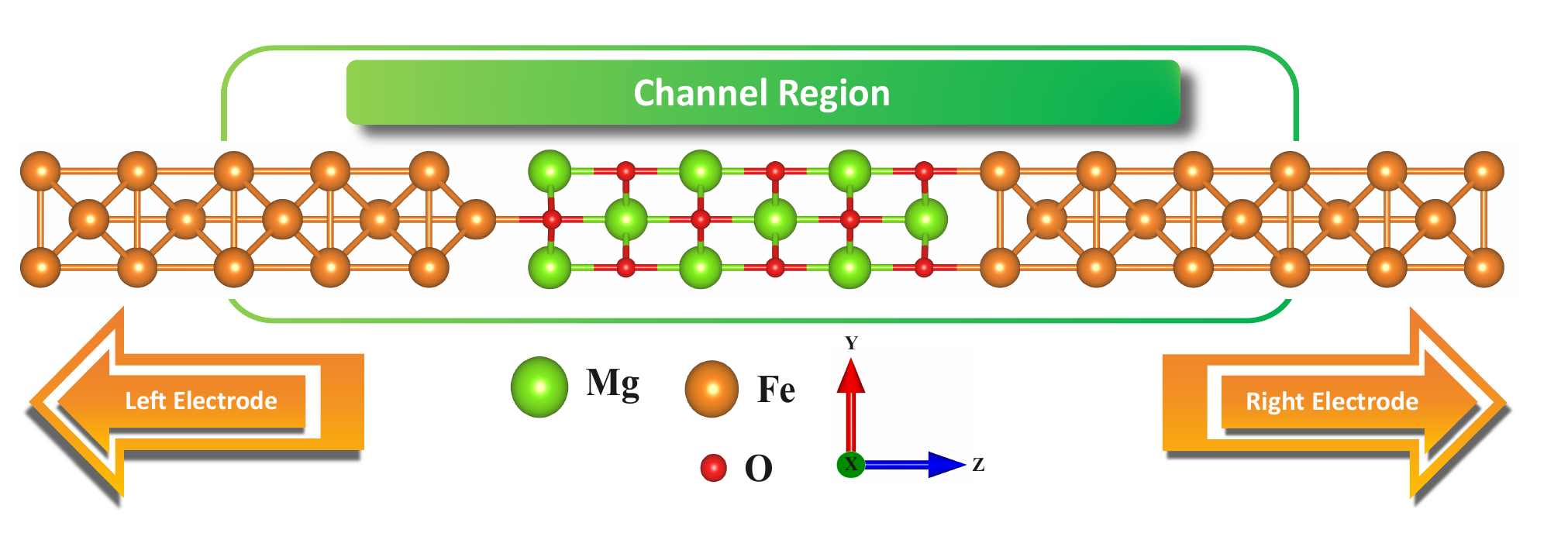} 
    \caption{Schematic of two probe Fe (100)/ MgO (100) /Fe (100) MTJ. The device extends with infinite periodicity in both the $x$ and the $y$ direction with a lattice constant of 2.847 $\text{\AA}$ and the Fe electrodes extend up to $\pm\infty$ to in the $z$-direction.}
    \label{fig:Device_MTJ}
\end{figure}
Rather than employing the Local Spin Density Approximation (LSDA) exchange-correlation, we opt for the GGA-PBE functional, which incorporates density gradient effects to better capture the impact of inhomogeneities of electron densities.
The distance between the Fe-O atoms at the interfaces is kept at 2.2$\AA$, which after the relaxation, turns out to be 2.228$\AA$, in agreement with earlier works\cite{butler2001spin,waldron2006first}. The Mermin free energy of the Fe/6-MgO/Fe structure (6l-MTJ) is found to be -1.49$\times 10 ^5$ eV, somewhat higher compared to the p-$\Delta_1^R$-MTJ(see Fig.~\ref{fig:FeO_distance}). 
 \begin{figure}[b]
    \centering
    \includegraphics[width= 8.2 cm]{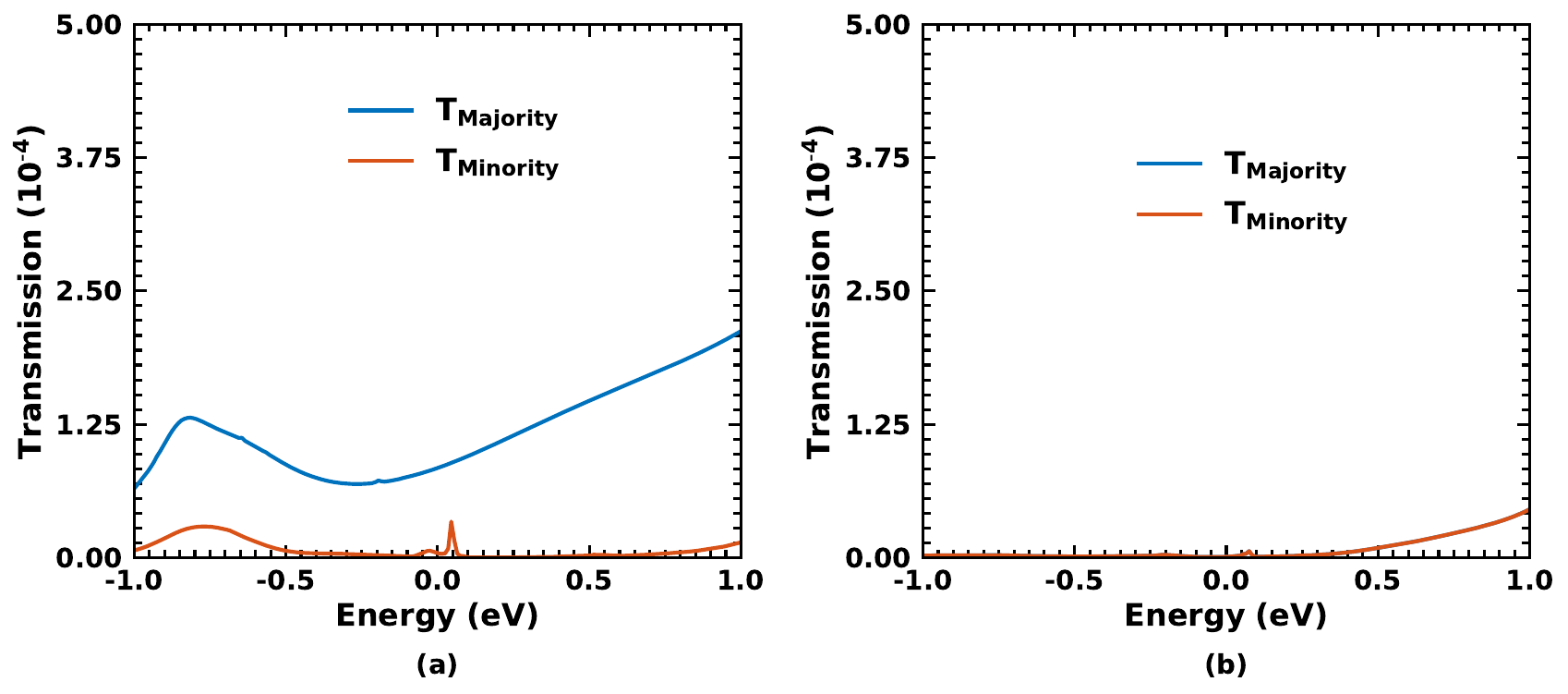} 
    \caption{The spin-resolved k-averaged transmission of the MTJ in the (a) PC and (b) the APC at zero bias.}
    \label{fig:k_avg_TM_MTJ}
\end{figure}
The k-averaged transmission, of the 6-layer(l) MTJ for the majority electrons, dominated by the $\Delta_1$ states in the PC, shows a smooth behaviour within the Fermi window, contrasted by that of the minority electrons with a sharp peak (see Fig.~\ref{fig:k_avg_TM_MTJ}(a)).
Both the majority and the minority electrons in the APC show a significantly diminished transmission(see Fig.~\ref{fig:k_avg_TM_MTJ}(b)) giving rise to a sizable TMR(\%) and spin polarisation of nearly 3400\% and 86\%, respectively.
The T$_{\text{Majority}}$ of the 6l-MTJ in the PC manifests a $\Gamma$-point centric signature(see Fig.~\ref{fig:k-res_MTJ}(a)), 
whereas T$_{\text{Minority}}$ displays a sharp interface resonance states at the Fermi energy(see Fig.~\ref{fig:k-res_MTJ}(b)). 
\begin{figure}[b]
    \centering
    \includegraphics[width= 8.5 cm]{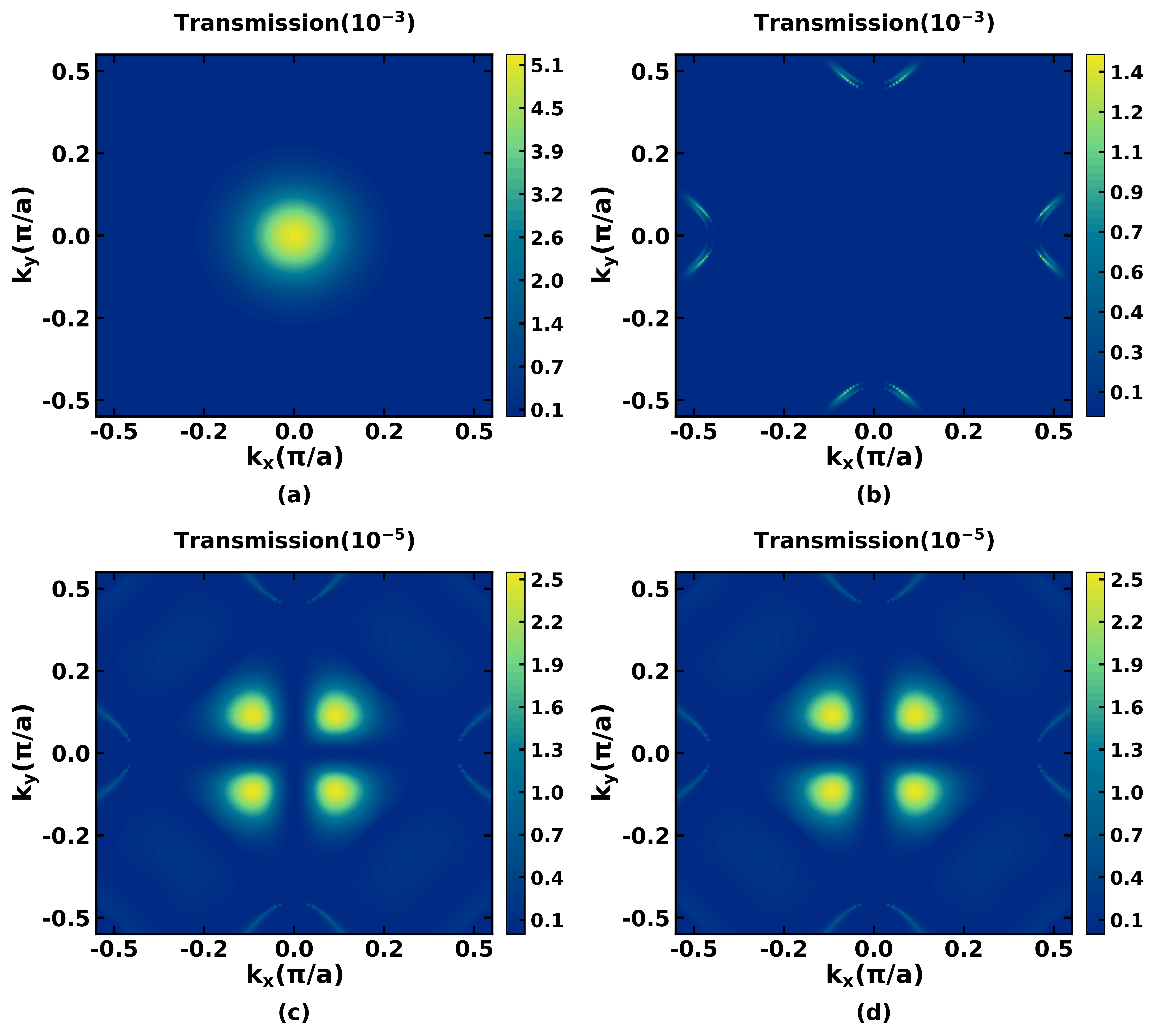} 
    \caption{k-resoved transmissions of the MTJ for the (a) majority and the (b) minority electrons in the PC, the (c) majority and the (d) minority electrons in the APC at Fermi energy with no applied bias.}
    \label{fig:k-res_MTJ}
\end{figure}
The T$_{\text{Majority}}$ and T$_{\text{Minority}}$ in the APC, presented a pattern, tracing a shape reminiscent of a flower with four radiant petals(see Fig.~\ref{fig:k-res_MTJ}(c),(d)), in agreement with previous studies\cite{waldron2006first}. 
\begin{figure}[t]
    \centering
    \includegraphics[width= 8.5 cm]{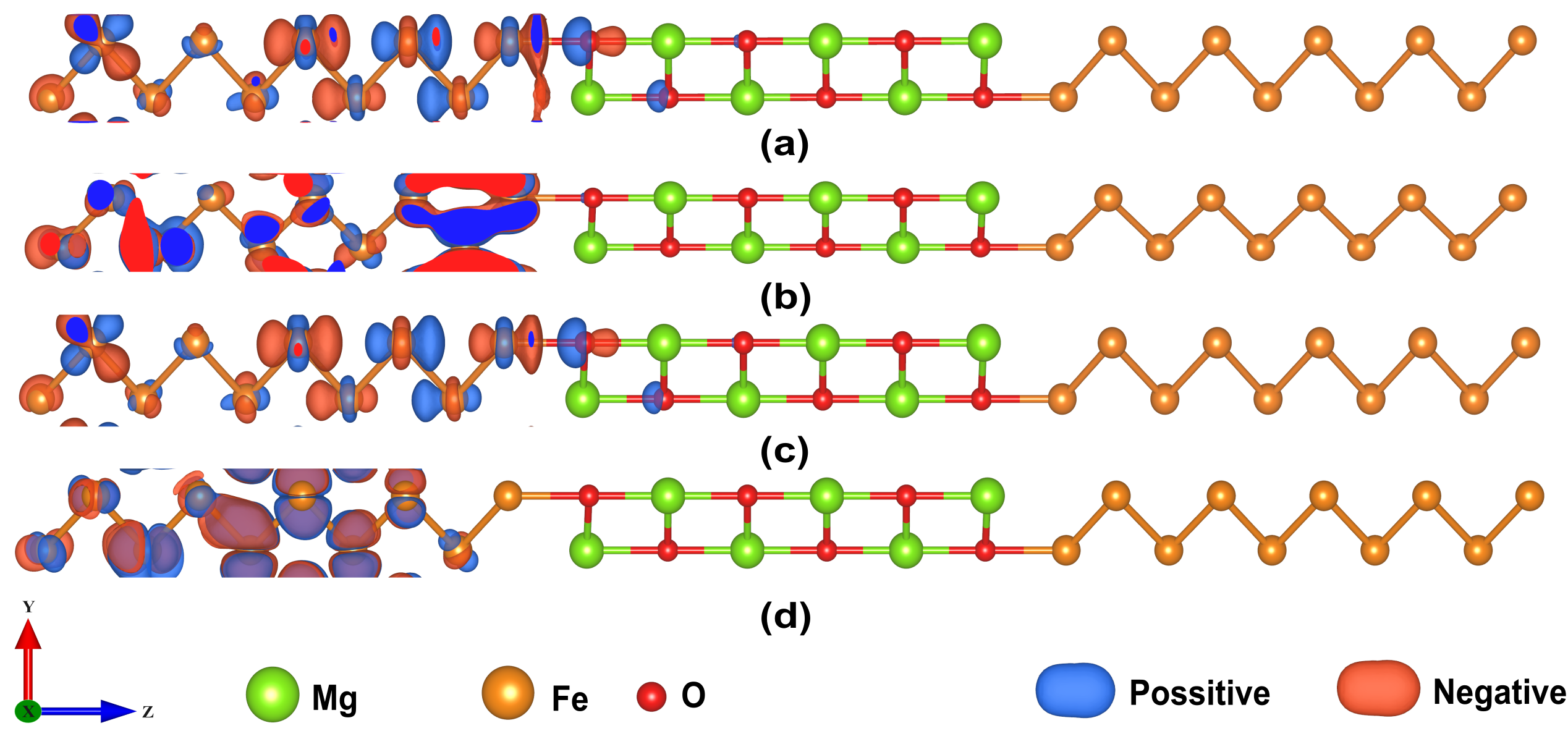} 
    \caption{c-eigenchannels of the $6l$-MTJ, emanating from the left electrode for the (a)majority and (b) minority electrons in the PC at $\lvert k_x \rvert = 0$, $\lvert k_y \rvert = 0$, followed by that of the (c) majority and the (d) minority electrons in the APC with $\lvert k_x \rvert = 0.12\pi$, $\lvert k_y \rvert = 0.12\pi$ at Fermi energy with no applied bias. The blue and red isosurfaces represent eigen wavefunctions with isovalues 0.1 and -0.1, respectively.}
    \label{fig:C_MTJ}
\end{figure}

We take a two-fold approach to probe into the transport features of the MTJ. In the first place, we present the eigenchannel and subsequently delve into the electrode-injected spectral density (SD) while corroborating our results with the orbital resolved bandstructure shown in  
Fig.~\ref{fig:BD_MTJ}. The band structures are plotted along the transport direction at a k-point, where the k-resolved transmission shown in Fig.~\ref{fig:k-res_MTJ} is found to have a considerable magnitude. 
In both the r-MTJ and the $\Delta_1^R$-MTJ, all the eigenchannels, except the one that we abbreviate as the conducting eigenchannel (c-eigenchannel), exhibit negligible transmissions and are referred to as the non-conducting eigenchannels (nc-eigenchannels).
\begin{figure}[b]
    \centering
    \includegraphics[width= 8.5 cm]{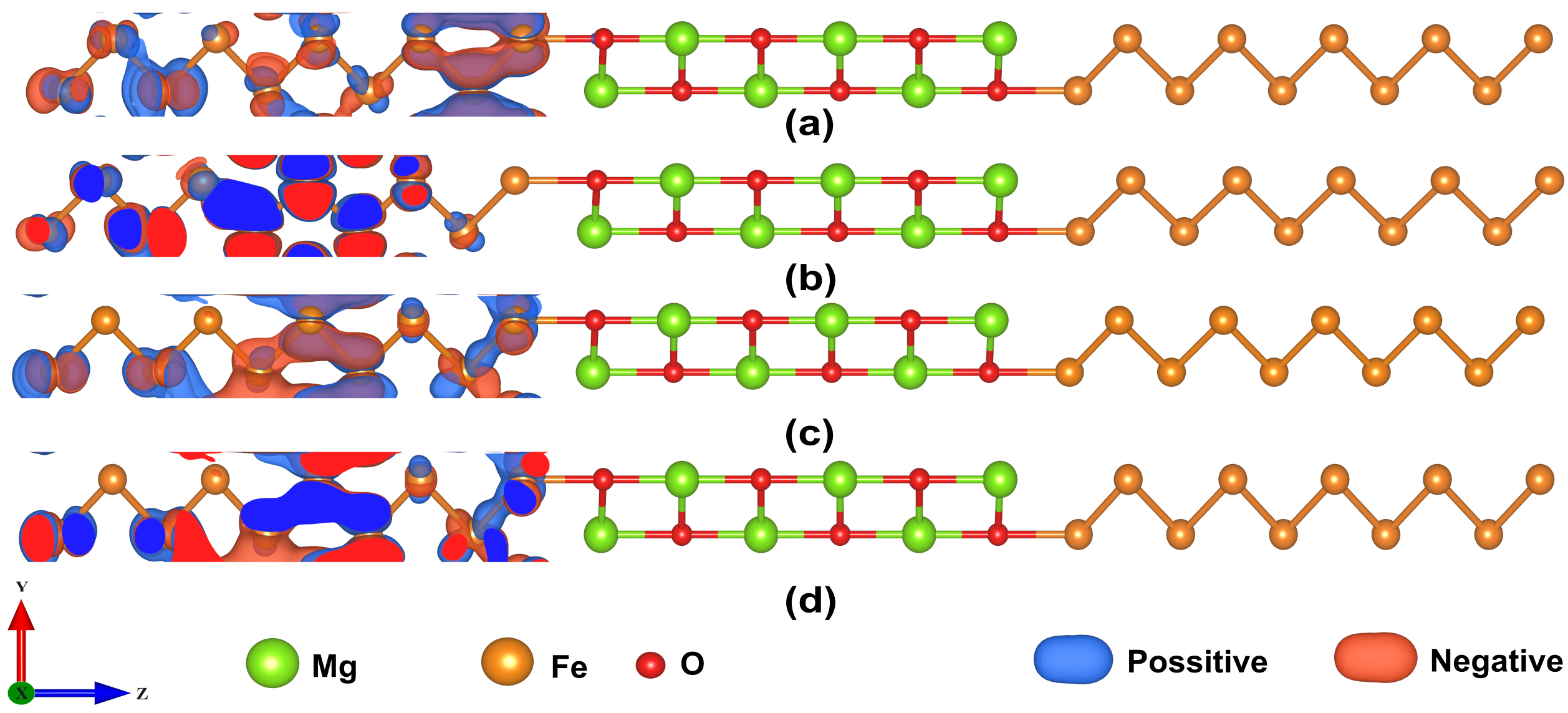} 
    \caption{nc-eigenchannels of the $6l$-MTJ, emanating from the left electrode for the (a)majority and (b) minority electrons in the PC at $\lvert k_x \rvert = 0$, $\lvert k_y \rvert = 0$, followed by that of the (c) majority and the (d) minority electrons in the APC with $\lvert k_x \rvert = 0.12\pi$, $\lvert k_y \rvert = 0.12\pi$ at Fermi energy with no applied bias. The blue and red isosurfaces represent transmission eigen-wavefunction with isovalues 0.1 and -0.1, respectively.}
     \label{fig:NC_MTJ}
\end{figure}
In Fig.~\ref{fig:C_MTJ}, we present the transmission eigen-wavefunction corresponding to the c-eigenchannels of the majority and minority electrons in the PC and the APC. A closer look in Fig.~\ref{fig:C_MTJ}(a) demonstrate the dominant contribution of $\Delta_1$ states associated with the majority electrons in the PC, endorsed via the $d_{Z^2}$ orbital of the Fe contact (near the Fe/MgO left interface) as presented in the bandstructure at Fermi energy (see Fig.~\ref{fig:BD_MTJ}(a)). However, unlike the p-$\Delta_1^R$-MTJ, the eigen-wavefunction in the r-MTJ shows feeble penetration in the MgO layers, signifying a regular tunneling of $\Delta_1$ states. The SD injected by the left contact for the respective electron states, as shown in Fig.~\ref{fig:ADOS_MTJ_6}(a), underscores the prevelence of the $\Delta_1$ states compared to their non-$\Delta_1$ counterparts during the transport.
\begin{figure*}[t]
    \centering
    \includegraphics[width= 17 cm]{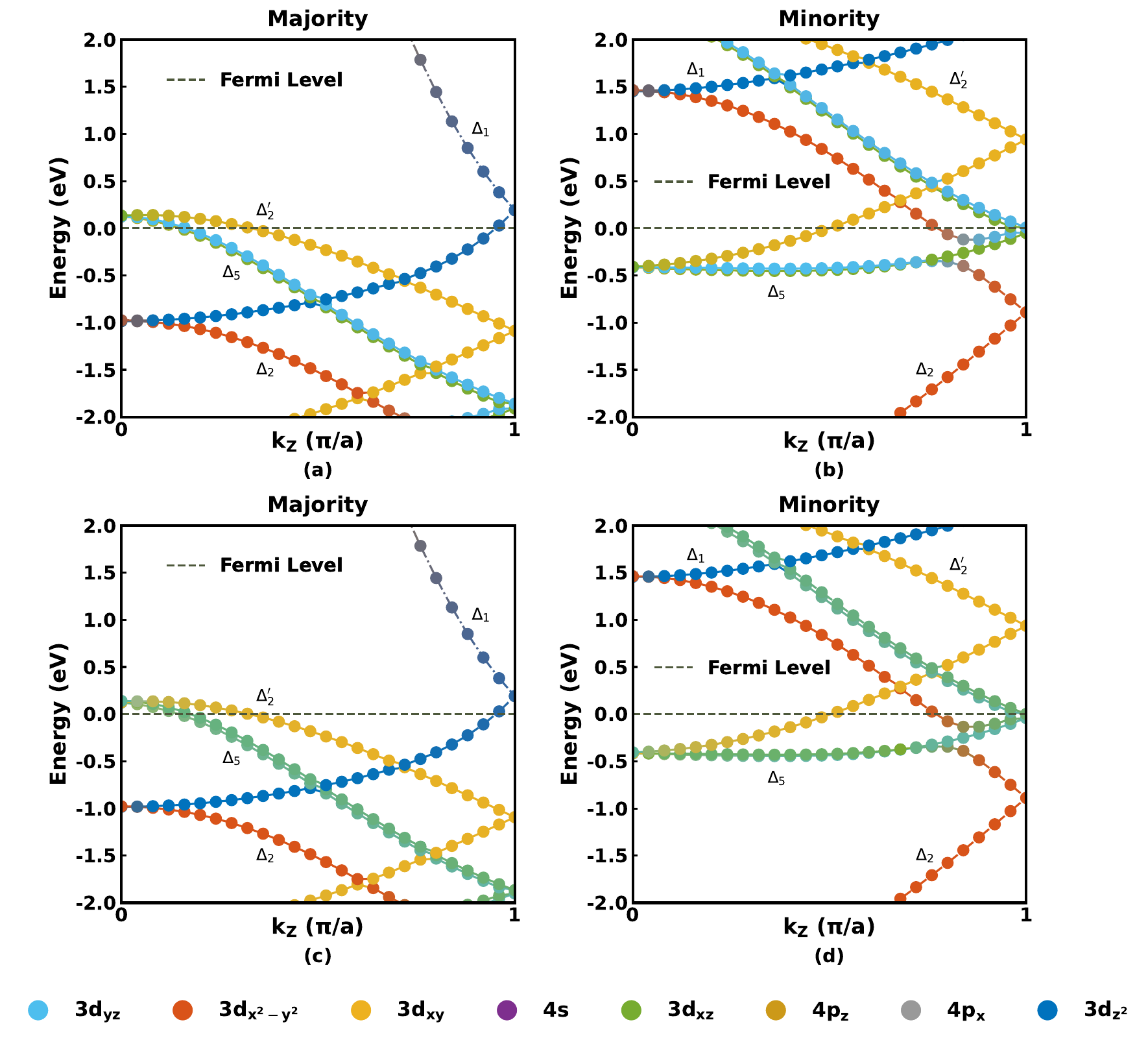} 
    \caption{The band structure of Fe electrodes are depicted for: (a) the majority and (b) the minority electrons at $\lvert k_x \rvert = 0$, $\lvert k_y \rvert = 0$; and (c)the majority and (d) the minority electrons at $\lvert k_x \rvert = 0.12\pi / a$, $\lvert k_y \rvert = 0.12\pi/a$, as $\lvert k_z \rvert$ is varied from $0$ to $\pi /a$. The colour coding of the bands corresponds to the orbital composition, as illustrated in the colour bar.}  
    \label{fig:BD_MTJ}
\end{figure*}
On the other hand, the c-eigenchannel of the minority electrons in the PC is shown in Fig.~\ref{fig:C_MTJ}(b), showcases the presence of the $d_{xz}$ and $d_{yz}$ orbitals, indicating the presence of the $\Delta_5$ states which show least decay for the minority states in the PC\cite{butler2001spin}. This observation is found to be consistent with Fig.~\ref{fig:BD_MTJ}(b) and also with Fig.~\ref{fig:ADOS_MTJ_6}(b), where we see a larger contribution of the non-$\Delta_1$ states in the SD. Similar to the $\Delta_1^R$-MTJ in the APC, presence of the $d_{Z^2}$ orbitals in the eigen-wavefunction shown Fig.~\ref{fig:C_MTJ}(c), highlights the coupling of the channel to the injecting left contact via the $\Delta_1$ states(see Fig.~\ref{fig:BD_MTJ}(c)), however, the decay of its spectral density depicted in Fig.~\ref{fig:ADOS_MTJ_6}(c), pointed out the absence of the $\Delta_1$ band in the terminal contact. The transport characteristics of minority electrons are notably more compelling contrasted to their majority counterparts, with the distinction that the $\Delta_1$ band is present in the receiving electrode instead of the injecting one. The eigenchannel of the minority electrons in the APC shown in Fig.~\ref{fig:C_MTJ}(d) shows the involvement of the $d_{xz}$, $d_{yz}$, $d_{x^2-y^2}$ and $d_{xy}$ orbitals, highlighting the presence of the  $\Delta_5$, $\Delta_2$ and $\Delta_2^{\prime}$ states. The SD of the $\Delta_1$ and the non-$\Delta_1$ states for them in the APC is shown in Fig.~\ref{fig:ADOS_MTJ_6}(d). The presence of the $\Delta_1$ states in the right electrode (majority states, see Fig.~\ref{fig:BD_MTJ}(c))) allows the injecting left contact to impart a higher spectral density compared to the non-$\Delta_1$ states. This, at first glance, feels rather unusual, as the minority bands of the injecting left contact near the Fermi level(see Fig.~\ref{fig:BD_MTJ}(d)) appeared to be free of the $\Delta_1$ character. Yet, analysing the fatbands reveals a nominal contribution of the $d_{Z^2}$ orbitals that barely impacts the colour composition. 
\begin{figure}[h]
    \centering
    \includegraphics[width= 8.5 cm]{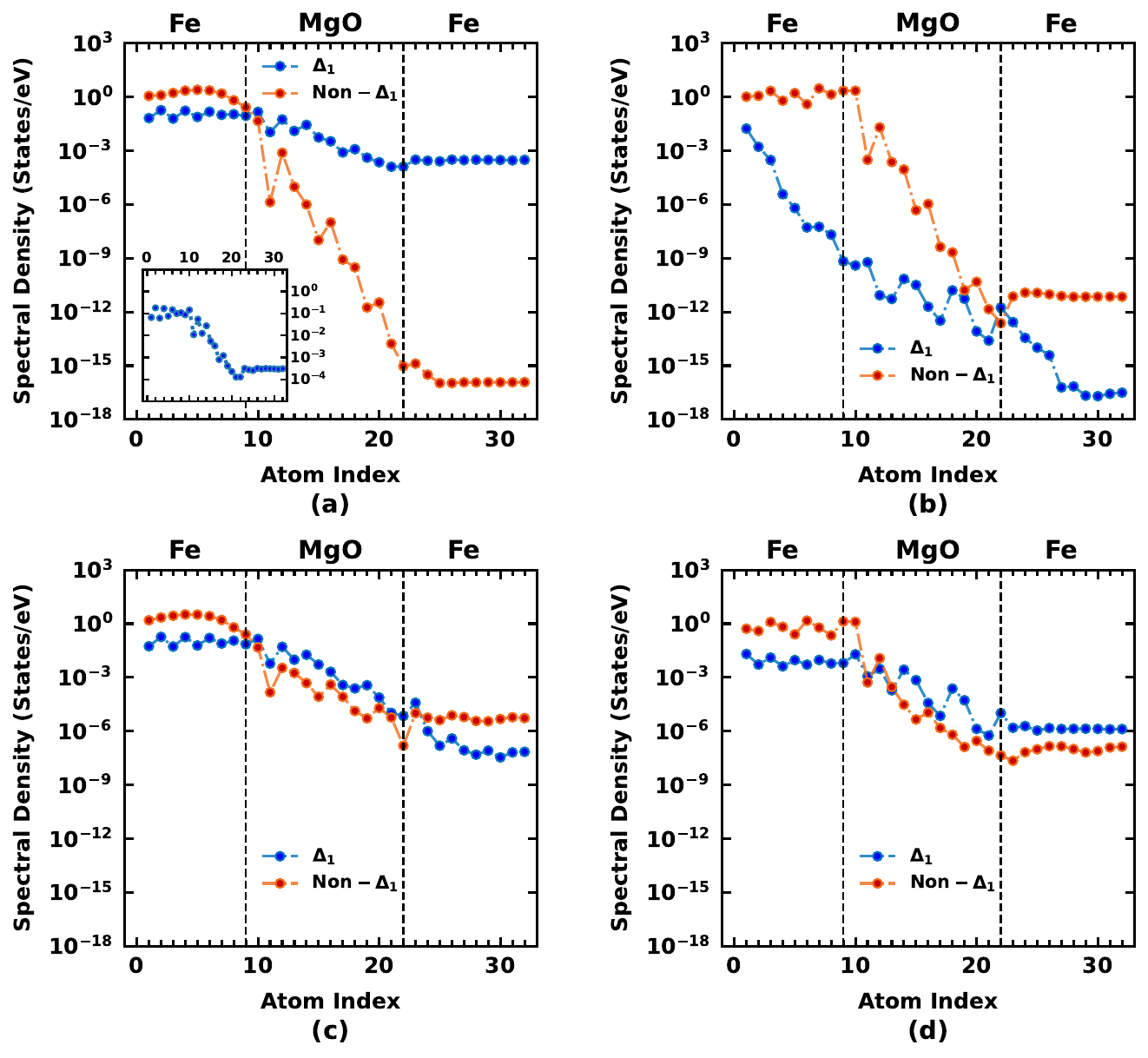} 
    \caption{Projected spectral density of the MTJ for the (a) majority and the (b) minority electrons originating from the left electrode in the PC at $\lvert k_x\lvert= 0$, $\lvert k_y \lvert = 0$ followed by that of the(c) majority and the (d) minority electrons in the APC with $\lvert k_x \lvert = 0.12\pi/a$, $\lvert k_y \lvert = 0.12\pi/a$ at Fermi energy.}
    \label{fig:ADOS_MTJ_6}
\end{figure}
The non-conducting eigenchannels for both the majority and minority electrons in the PC and the APC(see Fig.~\ref{fig:NC_MTJ}), shows the involvement of the $d_{xz}$, $d_{yz}$, $d_{x^2-y^2}$ and $d_{zx}$ orbitals, showcasing the presence of the $\Delta_2$, $\Delta_2^{\prime}$ and $\Delta_5$ states. Among all the nc-eigenchannels, those with the largest transmissions are at least five orders of magnitude lower than their conducting counterparts. Notably, none of the nc-eigenchannels exhibit any $\Delta_1$ character.

\section*{Appendix B: Non-conducting Eigenchannels of the $\text{p}$-$\Delta_1^R$-MTJ}
\label{appendix B}
\begin{figure}[b]
    \centering
    \includegraphics[width= 8.5 cm]{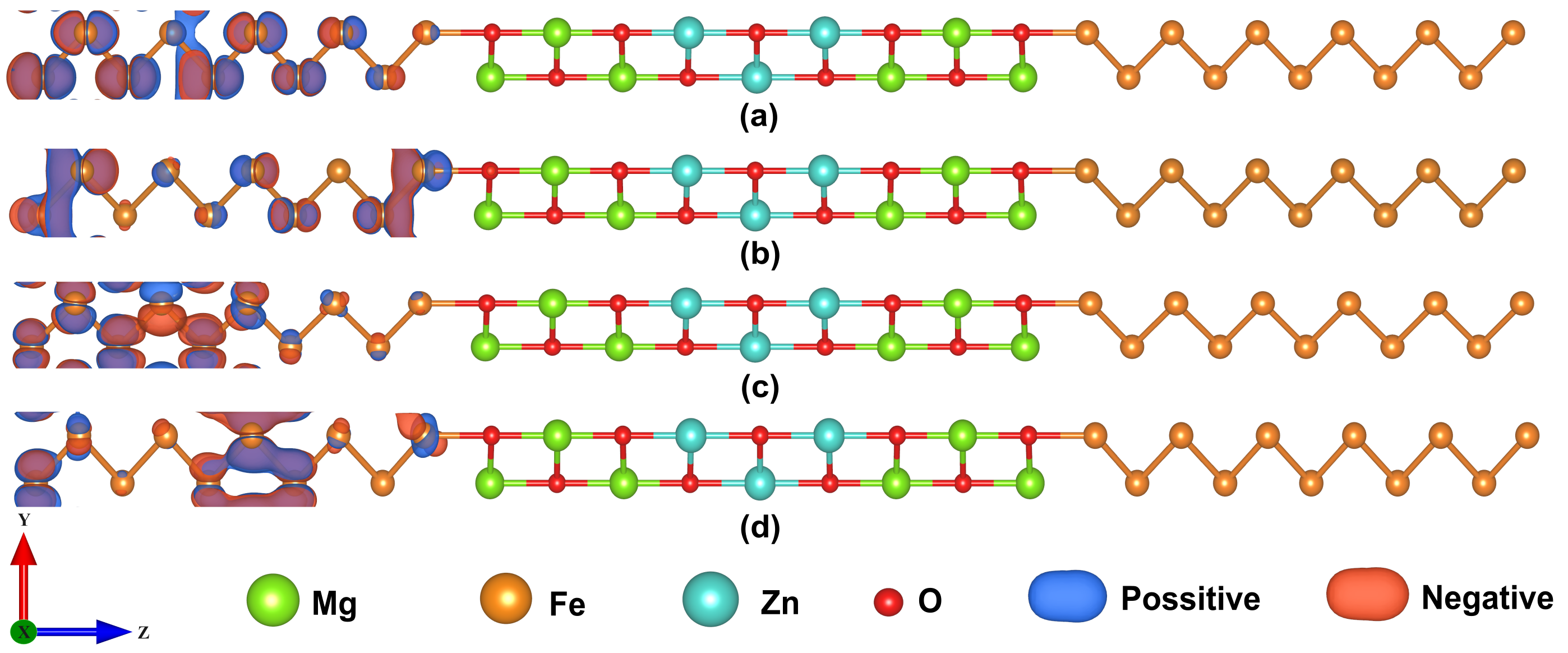} 
    \caption{nc-eigenchannels of the p-$\Delta^R_1$-MTJ for the (a) majority and the (c) minority electrons in the PC with $\lvert k_x \lvert = 0$, $\lvert k_y \lvert = 0.11\pi / a$, followed by the (c) majority and the (d) minority electrons in the APC at $\lvert k_x \lvert = 0.08\pi \lvert/ a$, $\lvert k_y \lvert = 0.06\pi/a$ at Fermi energy with no applied bias. The blue and red isosurfaces represent eigen wavefunctions with isovalues 0.1 and -0.1, respectively. In (b), the isosurface is given for a magnitude of 0.03.}
    \label{fig:NC_333}
\end{figure}

Similar to the r-MTJ, the nc-eigenchannels in the $\Delta_1^R$-MTJ highlights
the presence of the $d_{xz}$, $d_{yz}$, $d_{x^2-y^2}$, and $d_{xy}$ orbitals corroborating the involvement of the non-$\Delta_1$ states as shown in Fig.~\ref{fig:NC_333}.  Here as well, the nc-eigenchannels with the highest transmissions trailed their respective c-eigenchannels by over five orders of magnitude. Notably, none of the nc-eigenchannels exhibit any $\Delta_1$ character.
\section*{Appendix C: Stability of the $\Delta_1^R$-MTJ$\text{s}$}
\label{appendix C}
In finite-temperature density functional theory (DFT), the interatomic forces arise from minimizing the Mermin free energy\cite{mermin1965thermal}, which combines the internal energy and the electronic entropy. 
\begin{figure}[h]
    \centering
    \includegraphics[width= 8.5 cm]{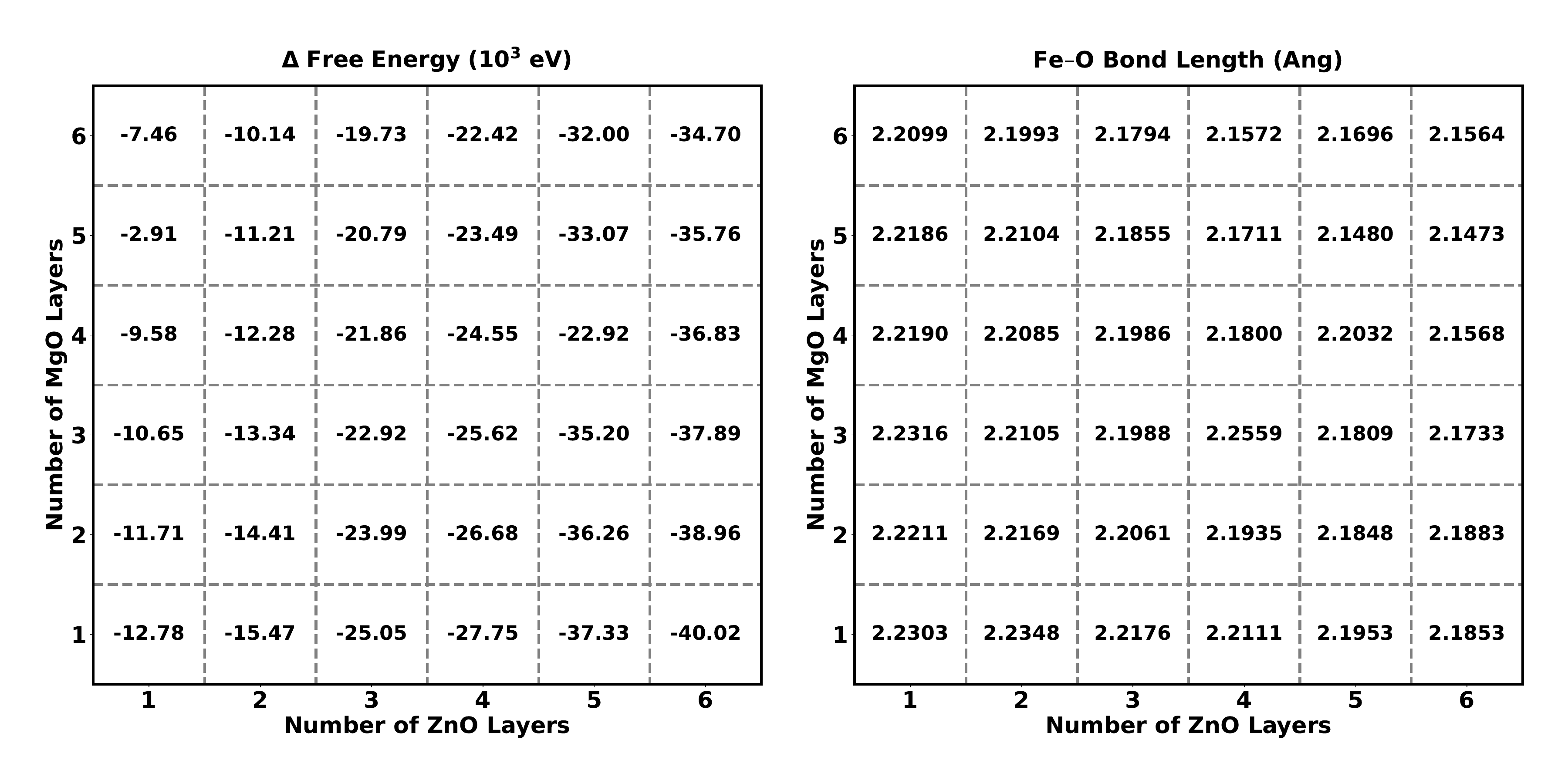} 
    \caption{(a) Stability of the $\Delta^R_1-MTJs$ are anticipated with respect to the $r-MTJ$ by subtracting the Mermin free energies of the devices.(b)The distance between the O atom in the terminal MgO layer and the Fe contact is reported with precision up to the fourth decimal digit. Here, an element with the nth row and mth column represents a $\Delta_1^R$ -MTJ featuring an n-MgO/m-ZnO/n-MgO sandwich flanked between the Fe layers.}
    \label{fig:FeO_distance}
\end{figure}
An \emph{electronic temperature} $T$ is introduced for metals or systems with small band gaps to smooth orbital occupancies, thus improving convergence in the self-consistent field (SCF) procedure. 
At this finite temperature, Kohn--Sham states are partially occupied according to the Fermi--Dirac function,
\[
f_i = \frac{1}{1 + e^{(\epsilon_i - \mu)/k_B T}},
\]
and the accompanying electronic entropy is given by
\[
S = - k_B \sum_i 
\bigl[
f_i \ln(f_i) + 
(1 - f_i)\ln\bigl(1 - f_i\bigr)
\bigr].
\]
The Mermin free energy, often represented as
\[
\Omega = U - T\,S,
\]
incorporates the Kohn-Sham potential energy $U$ and the entropy term $S$ arising from the partial occupancies. 
While the parameter $T$ is chosen to be large enough to smooth electronic occupancies near the Fermi level, it is typically kept small enough to avoid unphysical effects on the system's overall behaviour. 
The stability of the $\Delta_1^R$-MTJs is assessed relative to the r-MTJ (Fig.~\ref{fig:Device_MTJ}) by comparing their free energies, as it accounts for electronic entropy at finite temperature, unlike the Kohn–Sham energy\cite{wentzcovitch1992energy}. 
In Fig.~\ref{fig:FeO_distance}(a), we present the difference between the free energies between the $\Delta_1^R$-MTJs and the r-MTJ. The negative values of all the matrix elements highlights the boost in the stability of the $\Delta_1$-MTJs, likely facilitated by the ZnO quantum well, which contributes to a denser electron cloud around the Fe–O bond. This trend is also noticed in the table presented in Fig.~\ref{fig:FeO_distance}.
Although not monotonous, the introduction of the ZnO layers results in a reduction of the Fe-O bond length at the MgO-Fe interface. The value of the Fe-O bond length shows a generic trend of going up as the number of MgO layers increases. The devices with a higher percentage of ZnO layers have lower bond lengths. This is possibly due to the fact that increasing the number of MgO layers shielded the coupling of the Fe electrode with the energy states within the ZnO.

\section*{Appendix D: The redundancy of the GW calculation for majority states in the $\text{r}$-MTJ and the $\Delta_1^R$-MTJ}
\label{appendix D}
The requirement of GW calculation stems from the fact that LDA/GGA fails to capture the electron-electron correlations in insulating systems\cite{jiang2010first} such as MgO\cite{schonberger1995bulk}. However, metallic systems, with their nearly spherical Fermi surfaces, exhibit a strong screening effect, enabling LDA/GGA-PBE to predict the band structure of metals accurately\cite{haule2024all}. 
GW calculation performed on the Fe-MgO-Fe heterostructure reveals that the local DOS(LDOS) and transmission for the majority electrons show a negligible difference to that produced by LSDA/GGA-PBE calculations\cite{PhysRevB.110.224425, PhysRevB.85.174433}.  However, the GW correction, in agreement with experimental studies\cite{PhysRevB.78.033301, PhysRevB.81.064438}, yields a positive shift of 0.14–0.2 eV to the LDOS and transmission of the minority-spin interface resonance states \cite{PhysRevB.110.224425, PhysRevB.85.174433}.
\begin{figure}[h]
    \centering
    \includegraphics[width= 8.5 cm]{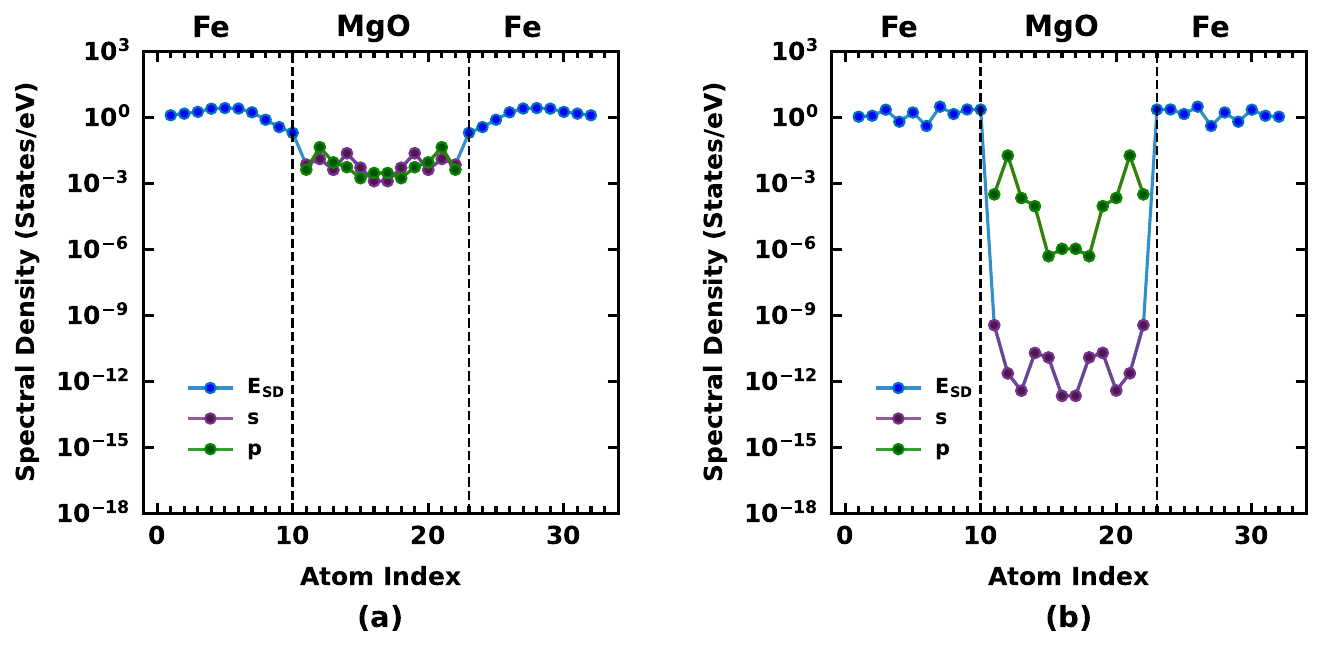} 
    \caption{(a) The $s$ and $p$ character of the spectral density for majority electrons, followed that of the of (b) minority electrons injected via the electrodes, evaluated at $\lvert k_x\rvert = 0$, $\lvert k_y\rvert = 0$, and the Fermi level in the PC for the r-MTJ. Here, E$_{SD}$ represents the total SD injected by both the contacts.}
    \label{fig:ADOS_MTJ_s}
\end{figure}
To seek a potential explanation, we examine the $s$ and $p$ characteristics of the electrode-injected SD associated with the MgO layers in the r-MTJ. Fig.~\ref{fig:ADOS_MTJ_s}(a) illustrates the dominant $s$-character (de-localized) of the majority electrons injected by Fe contacts through the $\Delta_1$ bands at the Fermi level, possibly promoting significant electron delocalization and shielding of electron-electron correlations. This may account for the agreement between the GW and GGA-PBE calculations for majority electrons in Fe-MgO-Fe heterostructure.
\begin{figure}[h]
    \centering
    \includegraphics[width= 8.5 cm]{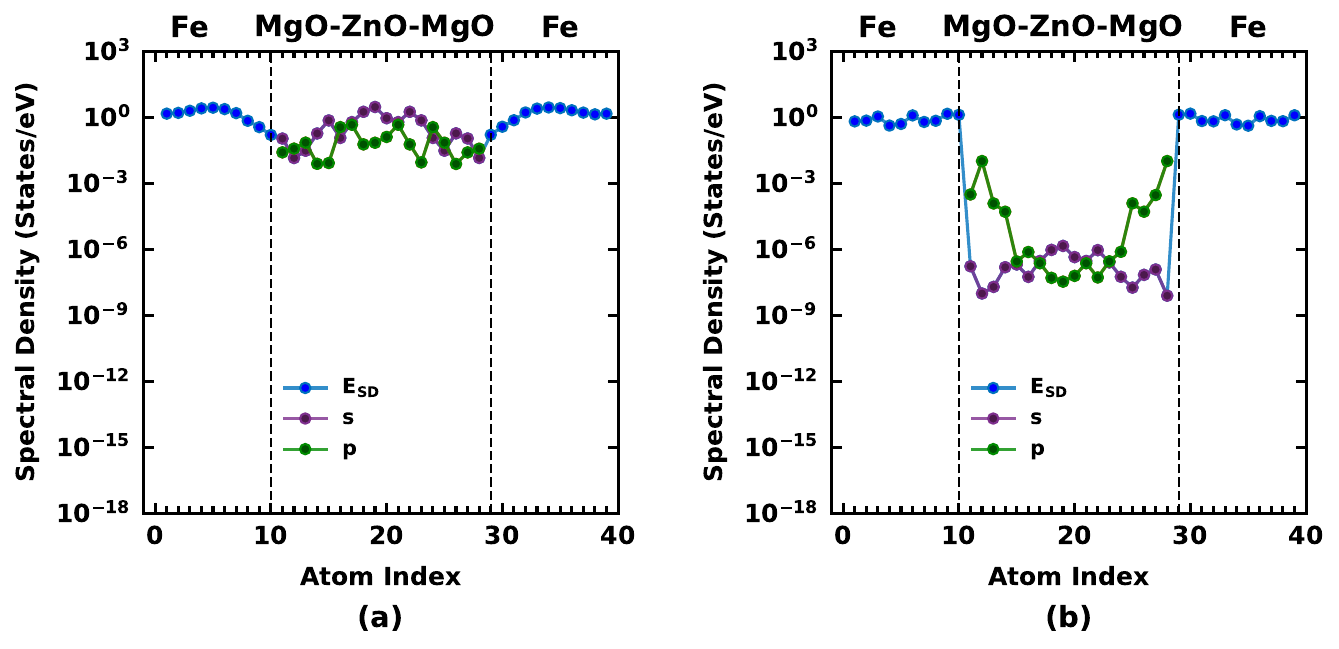} 
    \caption{(a) The $s$ and $p$ character of the spectral density for majority electrons, followed that of the of (b) minority electrons injected via the electrodes, evaluated at $\lvert k_x\rvert = 0.08 \pi/a$, $\lvert k_y\rvert = 0.06 \pi/a$, and the Fermi level in the PC for the $\Delta_1^R$-MTJ. Here, E$_{SD}$ represents the total SD injected by both the contacts.}
    \label{fig:ADOS_333_s}
\end{figure}
In contrast, the scenario for minority electrons depicted in Fig.~\ref{fig:ADOS_MTJ_s}(b) appears distinct, as the low concentration of $s$ electrons in over their $p$ counterparts in the MgO as layers likely leads to weaker delocalization and shielding. This results in a positive shift of approximately 0.14 eV in the minority spin LDOS and transmission between the GGA-PBE and the GW calculations\cite{PhysRevB.110.224425}. 

Likewise, the $\Delta_1^R$-MTJ with a relatively high $s$ character(see Fig.~\ref{fig:ADOS_333_s}(a)), as compared to the r-MTJ, is expected to show barely perceivable deviation in the transmission of majority electrons following the GW calculations owing to the symmetry selective resonance. The $s$ character of the SD showcased in Fig.~\ref{fig:ADOS_333_s}(a) is notably larger compared to its $p$ counterpart, unlike the r-MTJ(see Fig.~\ref{fig:ADOS_MTJ_s}(a)), indicating a higher delocalization. Consequently, the GW calculations are deemed even more redundant for majority states of $\Delta_1^R$-MTJ. On the other hand, the $s$ character of the minority electrons, as shown in Fig.~\ref{fig:ADOS_333_s}(a), is 3-4 orders higher than that of the r-MTJ, with the additional merit of the $s$ character nearly being the same as that of the $p$ character, implying an increase in the electron screening within the channel. Therefore, it is expected that the shift in the LDOS of the minority electrons upon GW calculation would be lower than the 0.14 eV observed in the r-MTJ. 
The reduction of the RA product in the PC and the increase in the TMR is primarily driven by the symmetry selective resonance of the majority electrons(see Fig. 5(a) in the main paper). The sizable $s$ character of the SD increases the screening significantly. We, therefore, expect the results to remain largely unaffected after GW calculation. As the GW calculation suppresses the ``hot spots" in the transverse momentum space for minority electrons in the PC and all electrons in the APC, it is likely to contribute to an increase in TMR\cite{PhysRevB.110.224425}.

\section*{Appendix E: Mathematical Framework}
\label{appendix E}
The current density ($J_C$) through the scattering region in the Landauer--B\"uttiker formalism given by\cite{PhysRevB.31.6207}:
\begin{equation}
\begin{split}
J_C = \frac{q}{\hbar} \sum_{\sigma} 
\int_{-\infty}^{\infty} \int_{\mathrm{BZ}} 
T_{\sigma}^{\,k_x,\,k_y}(E,V)\,\bigl[f_{1}(E) - f_{2}(E)\bigr]\\
\,\mathrm{d}k_x\,\mathrm{d}k_y\,\mathrm{d}E,
\end{split}
\end{equation}
where, $f_L(E)=1/(1+e{(E-E_{F_L})}/kT)$ and $f_R(E)=1/(1+e^{(E-E_{F_R})}/kT)$ are the Fermi distribution functions of the left 
and the right contact, respectively. Here, $V$ represents the applied voltage, $k$ indicates the Boltzmann constant, $e$ represents the electronic charge, and $T$ denotes the absolute temperature. $\sigma\equiv\downarrow,\uparrow$ indicates the spin index, $E_{F_L}$ and $E_{F_R}$ denotes the Fermi energies of the left and the right contact. The transmission $T_{\sigma}^{\,k_x,\,k_y}(E,V)$ using the NEGF framework is given by\cite{datta2005quantum}: 
\begin{equation}
\begin{split}
T_{\sigma}^{\,k_x,\,k_y}(E,V) = \mathrm{Tr} \Bigl[ \Gamma_{L,\sigma}^{\,k_x,\,k_y}(E,V)\, G_{\sigma}^{R}(k_x, k_y, E,V) \\
\Gamma_{R,\sigma}^{\,k_x,\,k_y}(E,V)\, G_{\sigma}^{A}(k_x, k_y, E,V) \Bigr],
\end{split}
\end{equation}
where, \(\Gamma_{L,\sigma}^{\,k_x,\,k_y}(E,V)\) and \(\Gamma_{R,\sigma}^{\,k_x,\,k_y}(E,V)\) are the broadening matrices of the left and the right contacts, respectively. \(G_{\sigma}^{R}(k_x, k_y, E,V)\) and \(G_{\sigma}^{A}(k_x, k_y, E,V)\) are the retarded and advanced Green's functions of the scattering region.  The broadening matrices in terms of the self-energy matrices of the left and the right contacts are given by\cite{datta2005quantum} 
\begin{equation}
\Gamma^{k_x, k_y}_{L/R,\sigma}(E, V) = i\left[ \Sigma^{k_x, k_y}_{L/R,\sigma}(E, V) - \left(\Sigma^{k_x, k_y}_{L/R,\sigma}(E, V)\right)^\dagger\right],
\end{equation}
The self-energies for the contacts are defined as
\begin{equation}
\Sigma_{L/R}^{\sigma}(k_x, k_y, E, V) = \tau_{L/R}^\dagger G_{L/R}^{\sigma}(k_x, k_y, E, V) \tau_{L/R},
\end{equation}
where, \(G_{L/R}^{\sigma}(k_x, k_y, E, V)\) is the Green's function of the left/right electrode, and \(\tau_{L/R}\) represents the coupling matrix between the scattering region and the contacts.\\

\bibliography{References}

\begin{thebibliography}{70}%
\makeatletter
\providecommand \@ifxundefined [1]{%
 \@ifx{#1\undefined}
}%
\providecommand \@ifnum [1]{%
 \ifnum #1\expandafter \@firstoftwo
 \else \expandafter \@secondoftwo
 \fi
}%
\providecommand \@ifx [1]{%
 \ifx #1\expandafter \@firstoftwo
 \else \expandafter \@secondoftwo
 \fi
}%
\providecommand \natexlab [1]{#1}%
\providecommand \enquote  [1]{``#1''}%
\providecommand \bibnamefont  [1]{#1}%
\providecommand \bibfnamefont [1]{#1}%
\providecommand \citenamefont [1]{#1}%
\providecommand \href@noop [0]{\@secondoftwo}%
\providecommand \href [0]{\begingroup \@sanitize@url \@href}%
\providecommand \@href[1]{\@@startlink{#1}\@@href}%
\providecommand \@@href[1]{\endgroup#1\@@endlink}%
\providecommand \@sanitize@url [0]{\catcode `\\12\catcode `\$12\catcode `\&12\catcode `\#12\catcode `\^12\catcode `\_12\catcode `\%12\relax}%
\providecommand \@@startlink[1]{}%
\providecommand \@@endlink[0]{}%
\providecommand \url  [0]{\begingroup\@sanitize@url \@url }%
\providecommand \@url [1]{\endgroup\@href {#1}{\urlprefix }}%
\providecommand \urlprefix  [0]{URL }%
\providecommand \Eprint [0]{\href }%
\providecommand \doibase [0]{https://doi.org/}%
\providecommand \selectlanguage [0]{\@gobble}%
\providecommand \bibinfo  [0]{\@secondoftwo}%
\providecommand \bibfield  [0]{\@secondoftwo}%
\providecommand \translation [1]{[#1]}%
\providecommand \BibitemOpen [0]{}%
\providecommand \bibitemStop [0]{}%
\providecommand \bibitemNoStop [0]{.\EOS\space}%
\providecommand \EOS [0]{\spacefactor3000\relax}%
\providecommand \BibitemShut  [1]{\csname bibitem#1\endcsname}%
\let\auto@bib@innerbib\@empty
\bibitem [{\citenamefont {Julliere}(1975)}]{JULLIERE1975225}%
  \BibitemOpen
  \bibfield  {author} {\bibinfo {author} {\bibfnamefont {M.}~\bibnamefont {Julliere}},\ }\bibfield  {title} {\bibinfo {title} {Tunneling between ferromagnetic films},\ }\href {https://doi.org/https://doi.org/10.1016/0375-9601(75)90174-7} {\bibfield  {journal} {\bibinfo  {journal} {Physics Letters A}\ }\textbf {\bibinfo {volume} {54}},\ \bibinfo {pages} {225} (\bibinfo {year} {1975})}\BibitemShut {NoStop}%
\bibitem [{\citenamefont {Butler}\ \emph {et~al.}(2001{\natexlab{a}})\citenamefont {Butler}, \citenamefont {Zhang}, \citenamefont {Schulthess},\ and\ \citenamefont {MacLaren}}]{PhysRevB.63.054416}%
  \BibitemOpen
  \bibfield  {author} {\bibinfo {author} {\bibfnamefont {W.~H.}\ \bibnamefont {Butler}}, \bibinfo {author} {\bibfnamefont {X.-G.}\ \bibnamefont {Zhang}}, \bibinfo {author} {\bibfnamefont {T.~C.}\ \bibnamefont {Schulthess}},\ and\ \bibinfo {author} {\bibfnamefont {J.~M.}\ \bibnamefont {MacLaren}},\ }\bibfield  {title} {\bibinfo {title} {Spin-dependent tunneling conductance of $\mathrm{Fe}|\mathrm{MgO}|\mathrm{Fe}$ sandwiches},\ }\href {https://doi.org/10.1103/PhysRevB.63.054416} {\bibfield  {journal} {\bibinfo  {journal} {Phys. Rev. B}\ }\textbf {\bibinfo {volume} {63}},\ \bibinfo {pages} {054416} (\bibinfo {year} {2001}{\natexlab{a}})}\BibitemShut {NoStop}%
\bibitem [{\citenamefont {Parkin}\ \emph {et~al.}(2004)\citenamefont {Parkin}, \citenamefont {Kaiser}, \citenamefont {Panchula}, \citenamefont {Rice}, \citenamefont {Hughes}, \citenamefont {Samant},\ and\ \citenamefont {Yang}}]{parkin2004giant}%
  \BibitemOpen
  \bibfield  {author} {\bibinfo {author} {\bibfnamefont {S.~S.}\ \bibnamefont {Parkin}}, \bibinfo {author} {\bibfnamefont {C.}~\bibnamefont {Kaiser}}, \bibinfo {author} {\bibfnamefont {A.}~\bibnamefont {Panchula}}, \bibinfo {author} {\bibfnamefont {P.~M.}\ \bibnamefont {Rice}}, \bibinfo {author} {\bibfnamefont {B.}~\bibnamefont {Hughes}}, \bibinfo {author} {\bibfnamefont {M.}~\bibnamefont {Samant}},\ and\ \bibinfo {author} {\bibfnamefont {S.-H.}\ \bibnamefont {Yang}},\ }\bibfield  {title} {\bibinfo {title} {Giant tunnelling magnetoresistance at room temperature with mgo (100) tunnel barriers},\ }\href@noop {} {\bibfield  {journal} {\bibinfo  {journal} {Nature materials}\ }\textbf {\bibinfo {volume} {3}},\ \bibinfo {pages} {862} (\bibinfo {year} {2004})}\BibitemShut {NoStop}%
\bibitem [{\citenamefont {Butler}\ \emph {et~al.}(2001{\natexlab{b}})\citenamefont {Butler}, \citenamefont {Zhang}, \citenamefont {Schulthess},\ and\ \citenamefont {MacLaren}}]{butler2001spin}%
  \BibitemOpen
  \bibfield  {author} {\bibinfo {author} {\bibfnamefont {W.}~\bibnamefont {Butler}}, \bibinfo {author} {\bibfnamefont {X.-G.}\ \bibnamefont {Zhang}}, \bibinfo {author} {\bibfnamefont {T.}~\bibnamefont {Schulthess}},\ and\ \bibinfo {author} {\bibfnamefont {J.}~\bibnamefont {MacLaren}},\ }\bibfield  {title} {\bibinfo {title} {Spin-dependent tunneling conductance of fe| mgo| fe sandwiches},\ }\href@noop {} {\bibfield  {journal} {\bibinfo  {journal} {Physical Review B}\ }\textbf {\bibinfo {volume} {63}},\ \bibinfo {pages} {054416} (\bibinfo {year} {2001}{\natexlab{b}})}\BibitemShut {NoStop}%
\bibitem [{\citenamefont {Waldron}\ \emph {et~al.}(2006)\citenamefont {Waldron}, \citenamefont {Timoshevskii}, \citenamefont {Hu}, \citenamefont {Xia},\ and\ \citenamefont {Guo}}]{waldron2006first}%
  \BibitemOpen
  \bibfield  {author} {\bibinfo {author} {\bibfnamefont {D.}~\bibnamefont {Waldron}}, \bibinfo {author} {\bibfnamefont {V.}~\bibnamefont {Timoshevskii}}, \bibinfo {author} {\bibfnamefont {Y.}~\bibnamefont {Hu}}, \bibinfo {author} {\bibfnamefont {K.}~\bibnamefont {Xia}},\ and\ \bibinfo {author} {\bibfnamefont {H.}~\bibnamefont {Guo}},\ }\bibfield  {title} {\bibinfo {title} {First principles modeling of tunnel magnetoresistance of fe/mgo/fe trilayers},\ }\href {https://doi.org/10.1103/PhysRevLett.97.226802} {\bibfield  {journal} {\bibinfo  {journal} {Physical review letters}\ }\textbf {\bibinfo {volume} {97}},\ \bibinfo {pages} {226802} (\bibinfo {year} {2006})}\BibitemShut {NoStop}%
\bibitem [{\citenamefont {Berger}(1996)}]{berger1996emission}%
  \BibitemOpen
  \bibfield  {author} {\bibinfo {author} {\bibfnamefont {L.}~\bibnamefont {Berger}},\ }\bibfield  {title} {\bibinfo {title} {Emission of spin waves by a magnetic multilayer traversed by a current},\ }\href {https://doi.org/https://doi.org/10.1103/PhysRevB.54.9353} {\bibfield  {journal} {\bibinfo  {journal} {Physical Review B}\ }\textbf {\bibinfo {volume} {54}},\ \bibinfo {pages} {9353} (\bibinfo {year} {1996})}\BibitemShut {NoStop}%
\bibitem [{\citenamefont {Slonczewski}(1996)}]{slonczewski1996current}%
  \BibitemOpen
  \bibfield  {author} {\bibinfo {author} {\bibfnamefont {J.~C.}\ \bibnamefont {Slonczewski}},\ }\bibfield  {title} {\bibinfo {title} {Current-driven excitation of magnetic multilayers},\ }\href@noop {} {\bibfield  {journal} {\bibinfo  {journal} {Journal of Magnetism and Magnetic Materials}\ }\textbf {\bibinfo {volume} {159}},\ \bibinfo {pages} {L1} (\bibinfo {year} {1996})}\BibitemShut {NoStop}%
\bibitem [{\citenamefont {Ouyang}\ \emph {et~al.}(2015)\citenamefont {Ouyang}, \citenamefont {He}, \citenamefont {Hu}, \citenamefont {Zhao}, \citenamefont {Wang},\ and\ \citenamefont {Wang}}]{7124469}%
  \BibitemOpen
  \bibfield  {author} {\bibinfo {author} {\bibfnamefont {Y.}~\bibnamefont {Ouyang}}, \bibinfo {author} {\bibfnamefont {J.}~\bibnamefont {He}}, \bibinfo {author} {\bibfnamefont {J.}~\bibnamefont {Hu}}, \bibinfo {author} {\bibfnamefont {G.}~\bibnamefont {Zhao}}, \bibinfo {author} {\bibfnamefont {Z.}~\bibnamefont {Wang}},\ and\ \bibinfo {author} {\bibfnamefont {S.~X.}\ \bibnamefont {Wang}},\ }\bibfield  {title} {\bibinfo {title} {Contactless current sensors based on magnetic tunnel junction for smart grid applications},\ }\href {https://doi.org/10.1109/TMAG.2015.2446332} {\bibfield  {journal} {\bibinfo  {journal} {IEEE Transactions on Magnetics}\ }\textbf {\bibinfo {volume} {51}},\ \bibinfo {pages} {1} (\bibinfo {year} {2015})}\BibitemShut {NoStop}%
\bibitem [{\citenamefont {Zink}\ \emph {et~al.}(2022)\citenamefont {Zink}, \citenamefont {Lv},\ and\ \citenamefont {Wang}}]{9976889}%
  \BibitemOpen
  \bibfield  {author} {\bibinfo {author} {\bibfnamefont {B.~R.}\ \bibnamefont {Zink}}, \bibinfo {author} {\bibfnamefont {Y.}~\bibnamefont {Lv}},\ and\ \bibinfo {author} {\bibfnamefont {J.-P.}\ \bibnamefont {Wang}},\ }\bibfield  {title} {\bibinfo {title} {Review of magnetic tunnel junctions for stochastic computing},\ }\href {https://doi.org/10.1109/JXCDC.2022.3227062} {\bibfield  {journal} {\bibinfo  {journal} {IEEE Journal on Exploratory Solid-State Computational Devices and Circuits}\ }\textbf {\bibinfo {volume} {8}},\ \bibinfo {pages} {173} (\bibinfo {year} {2022})}\BibitemShut {NoStop}%
\bibitem [{\citenamefont {Sun}\ \emph {et~al.}(2017)\citenamefont {Sun}, \citenamefont {Zhao}, \citenamefont {Poremba}, \citenamefont {Xu},\ and\ \citenamefont {Xie}}]{10.1093/nsr/nwx082}%
  \BibitemOpen
  \bibfield  {author} {\bibinfo {author} {\bibfnamefont {G.}~\bibnamefont {Sun}}, \bibinfo {author} {\bibfnamefont {J.}~\bibnamefont {Zhao}}, \bibinfo {author} {\bibfnamefont {M.}~\bibnamefont {Poremba}}, \bibinfo {author} {\bibfnamefont {C.}~\bibnamefont {Xu}},\ and\ \bibinfo {author} {\bibfnamefont {Y.}~\bibnamefont {Xie}},\ }\bibfield  {title} {\bibinfo {title} {{Memory that never forgets: emerging nonvolatile memory and the implication for architecture design}},\ }\href {https://doi.org/10.1093/nsr/nwx082} {\bibfield  {journal} {\bibinfo  {journal} {National Science Review}\ }\textbf {\bibinfo {volume} {5}},\ \bibinfo {pages} {577} (\bibinfo {year} {2017})}\BibitemShut {NoStop}%
\bibitem [{\citenamefont {Ielmini}\ and\ \citenamefont {Ambrogio}(2019)}]{Ielmini_2020}%
  \BibitemOpen
  \bibfield  {author} {\bibinfo {author} {\bibfnamefont {D.}~\bibnamefont {Ielmini}}\ and\ \bibinfo {author} {\bibfnamefont {S.}~\bibnamefont {Ambrogio}},\ }\bibfield  {title} {\bibinfo {title} {Emerging neuromorphic devices},\ }\href {https://doi.org/10.1088/1361-6528/ab554b} {\bibfield  {journal} {\bibinfo  {journal} {Nanotechnology}\ }\textbf {\bibinfo {volume} {31}},\ \bibinfo {pages} {092001} (\bibinfo {year} {2019})}\BibitemShut {NoStop}%
\bibitem [{\citenamefont {Grollier}\ \emph {et~al.}(2020)\citenamefont {Grollier}, \citenamefont {Querlioz}, \citenamefont {Camsari}, \citenamefont {Everschor-Sitte}, \citenamefont {Fukami},\ and\ \citenamefont {Stiles}}]{grollier2020neuromorphic}%
  \BibitemOpen
  \bibfield  {author} {\bibinfo {author} {\bibfnamefont {J.}~\bibnamefont {Grollier}}, \bibinfo {author} {\bibfnamefont {D.}~\bibnamefont {Querlioz}}, \bibinfo {author} {\bibfnamefont {K.~Y.}\ \bibnamefont {Camsari}}, \bibinfo {author} {\bibfnamefont {K.}~\bibnamefont {Everschor-Sitte}}, \bibinfo {author} {\bibfnamefont {S.}~\bibnamefont {Fukami}},\ and\ \bibinfo {author} {\bibfnamefont {M.~D.}\ \bibnamefont {Stiles}},\ }\bibfield  {title} {\bibinfo {title} {Neuromorphic computing with spintronics},\ }\href {https://doi.org/10.1038/s41928-020-0461-5} {\bibfield  {journal} {\bibinfo  {journal} {Nature Electronics}\ }\textbf {\bibinfo {volume} {3}},\ \bibinfo {pages} {360} (\bibinfo {year} {2020})}\BibitemShut {NoStop}%
\bibitem [{\citenamefont {Zhang}\ \emph {et~al.}(2023)\citenamefont {Zhang}, \citenamefont {Wang},\ and\ \citenamefont {Zhao}}]{zhang2023spintronic}%
  \BibitemOpen
  \bibfield  {author} {\bibinfo {author} {\bibfnamefont {X.}~\bibnamefont {Zhang}}, \bibinfo {author} {\bibfnamefont {J.}~\bibnamefont {Wang}},\ and\ \bibinfo {author} {\bibfnamefont {W.}~\bibnamefont {Zhao}},\ }\bibfield  {title} {\bibinfo {title} {Spintronic leaky-integrate-fire spiking neurons with self-reset and lateral inhibition},\ }\href {https://doi.org/10.1038/s41467-023-36728-1} {\bibfield  {journal} {\bibinfo  {journal} {Nature Communications}\ }\textbf {\bibinfo {volume} {14}},\ \bibinfo {pages} {36728} (\bibinfo {year} {2023})}\BibitemShut {NoStop}%
\bibitem [{\citenamefont {Zhang}\ \emph {et~al.}(2024)\citenamefont {Zhang}, \citenamefont {Huang}, \citenamefont {Li} \emph {et~al.}}]{zhang2024domain}%
  \BibitemOpen
  \bibfield  {author} {\bibinfo {author} {\bibfnamefont {X.}~\bibnamefont {Zhang}}, \bibinfo {author} {\bibfnamefont {Y.}~\bibnamefont {Huang}}, \bibinfo {author} {\bibfnamefont {H.}~\bibnamefont {Li}}, \emph {et~al.},\ }\bibfield  {title} {\bibinfo {title} {Domain wall-magnetic tunnel junction-based artificial synapses and neurons for neuromorphic computing},\ }\href {https://doi.org/10.1038/s41467-024-48631-4} {\bibfield  {journal} {\bibinfo  {journal} {Nature Communications}\ }\textbf {\bibinfo {volume} {15}},\ \bibinfo {pages} {48631} (\bibinfo {year} {2024})}\BibitemShut {NoStop}%
\bibitem [{\citenamefont {Camsari}\ \emph {et~al.}(2019)\citenamefont {Camsari}, \citenamefont {Sutton},\ and\ \citenamefont {Datta}}]{camsari2019pbits}%
  \BibitemOpen
  \bibfield  {author} {\bibinfo {author} {\bibfnamefont {K.~Y.}\ \bibnamefont {Camsari}}, \bibinfo {author} {\bibfnamefont {B.~M.}\ \bibnamefont {Sutton}},\ and\ \bibinfo {author} {\bibfnamefont {S.}~\bibnamefont {Datta}},\ }\bibfield  {title} {\bibinfo {title} {p-bits for probabilistic spin logic},\ }\href {https://doi.org/10.1109/TED.2019.2938710} {\bibfield  {journal} {\bibinfo  {journal} {IEEE Transactions on Electron Devices}\ }\textbf {\bibinfo {volume} {66}},\ \bibinfo {pages} {5023} (\bibinfo {year} {2019})}\BibitemShut {NoStop}%
\bibitem [{\citenamefont {Gao}\ \emph {et~al.}(2020)\citenamefont {Gao}, \citenamefont {Yin}, \citenamefont {Yang}, \citenamefont {Ewing}, \citenamefont {De~Rego},\ and\ \citenamefont {Liou}}]{10.1063/1.5129902}%
  \BibitemOpen
  \bibfield  {author} {\bibinfo {author} {\bibfnamefont {K.-Z.}\ \bibnamefont {Gao}}, \bibinfo {author} {\bibfnamefont {X.}~\bibnamefont {Yin}}, \bibinfo {author} {\bibfnamefont {Y.}~\bibnamefont {Yang}}, \bibinfo {author} {\bibfnamefont {D.}~\bibnamefont {Ewing}}, \bibinfo {author} {\bibfnamefont {P.~J.}\ \bibnamefont {De~Rego}},\ and\ \bibinfo {author} {\bibfnamefont {S.-H.}\ \bibnamefont {Liou}},\ }\bibfield  {title} {\bibinfo {title} {{MTJ based magnetic sensor for current measurement in grid}},\ }\href {https://doi.org/10.1063/1.5129902} {\bibfield  {journal} {\bibinfo  {journal} {AIP Advances}\ }\textbf {\bibinfo {volume} {10}},\ \bibinfo {pages} {015301} (\bibinfo {year} {2020})}\BibitemShut {NoStop}%
\bibitem [{\citenamefont {Cardoso}\ \emph {et~al.}(2014)\citenamefont {Cardoso}, \citenamefont {Leitao}, \citenamefont {Gameiro}, \citenamefont {Cardoso}, \citenamefont {Ferreira}, \citenamefont {Paz},\ and\ \citenamefont {Freitas}}]{cardoso2014magnetic}%
  \BibitemOpen
  \bibfield  {author} {\bibinfo {author} {\bibfnamefont {S.}~\bibnamefont {Cardoso}}, \bibinfo {author} {\bibfnamefont {D.}~\bibnamefont {Leitao}}, \bibinfo {author} {\bibfnamefont {L.}~\bibnamefont {Gameiro}}, \bibinfo {author} {\bibfnamefont {F.}~\bibnamefont {Cardoso}}, \bibinfo {author} {\bibfnamefont {R.}~\bibnamefont {Ferreira}}, \bibinfo {author} {\bibfnamefont {E.}~\bibnamefont {Paz}},\ and\ \bibinfo {author} {\bibfnamefont {P.}~\bibnamefont {Freitas}},\ }\bibfield  {title} {\bibinfo {title} {Magnetic tunnel junction sensors with ptesla sensitivity},\ }\href {https://doi.org/https://doi.org/10.1007/s00542-013-2035-1} {\bibfield  {journal} {\bibinfo  {journal} {Microsystem technologies}\ }\textbf {\bibinfo {volume} {20}},\ \bibinfo {pages} {793} (\bibinfo {year} {2014})}\BibitemShut {NoStop}%
\bibitem [{\citenamefont {Liu}\ \emph {et~al.}(2024)\citenamefont {Liu}, \citenamefont {Wang}, \citenamefont {Wang}, \citenamefont {Sun}, \citenamefont {Lin}, \citenamefont {Gong}, \citenamefont {Zhang}, \citenamefont {Tang}, \citenamefont {Mai}, \citenamefont {Hou} \emph {et~al.}}]{liu2024domain}%
  \BibitemOpen
  \bibfield  {author} {\bibinfo {author} {\bibfnamefont {L.}~\bibnamefont {Liu}}, \bibinfo {author} {\bibfnamefont {D.}~\bibnamefont {Wang}}, \bibinfo {author} {\bibfnamefont {D.}~\bibnamefont {Wang}}, \bibinfo {author} {\bibfnamefont {Y.}~\bibnamefont {Sun}}, \bibinfo {author} {\bibfnamefont {H.}~\bibnamefont {Lin}}, \bibinfo {author} {\bibfnamefont {X.}~\bibnamefont {Gong}}, \bibinfo {author} {\bibfnamefont {Y.}~\bibnamefont {Zhang}}, \bibinfo {author} {\bibfnamefont {R.}~\bibnamefont {Tang}}, \bibinfo {author} {\bibfnamefont {Z.}~\bibnamefont {Mai}}, \bibinfo {author} {\bibfnamefont {Z.}~\bibnamefont {Hou}}, \emph {et~al.},\ }\bibfield  {title} {\bibinfo {title} {Domain wall magnetic tunnel junction-based artificial synapses and neurons for all-spin neuromorphic hardware},\ }\href {https://doi.org/https://doi.org/10.1038/s41467-024-48631-4} {\bibfield  {journal} {\bibinfo  {journal} {Nature Communications}\ }\textbf {\bibinfo {volume} {15}},\ \bibinfo {pages} {4534} (\bibinfo {year} {2024})}\BibitemShut
  {NoStop}%
\bibitem [{\citenamefont {Joshi}\ \emph {et~al.}(2020)\citenamefont {Joshi}, \citenamefont {Barla}, \citenamefont {Bhat},\ and\ \citenamefont {Kaushik}}]{9235579}%
  \BibitemOpen
  \bibfield  {author} {\bibinfo {author} {\bibfnamefont {V.~K.}\ \bibnamefont {Joshi}}, \bibinfo {author} {\bibfnamefont {P.}~\bibnamefont {Barla}}, \bibinfo {author} {\bibfnamefont {S.}~\bibnamefont {Bhat}},\ and\ \bibinfo {author} {\bibfnamefont {B.~K.}\ \bibnamefont {Kaushik}},\ }\bibfield  {title} {\bibinfo {title} {From mtj device to hybrid cmos/mtj circuits: A review},\ }\href {https://doi.org/10.1109/ACCESS.2020.3033023} {\bibfield  {journal} {\bibinfo  {journal} {IEEE Access}\ }\textbf {\bibinfo {volume} {8}},\ \bibinfo {pages} {194105} (\bibinfo {year} {2020})}\BibitemShut {NoStop}%
\bibitem [{\citenamefont {Sharma}\ \emph {et~al.}(2024)\citenamefont {Sharma}, \citenamefont {Ngo}, \citenamefont {Raimondo}, \citenamefont {Giordano}, \citenamefont {Igarashi}, \citenamefont {Jinnai}, \citenamefont {Zhao}, \citenamefont {Lei}, \citenamefont {Guo}, \citenamefont {Finocchio} \emph {et~al.}}]{sharma2024nanoscale}%
  \BibitemOpen
  \bibfield  {author} {\bibinfo {author} {\bibfnamefont {R.}~\bibnamefont {Sharma}}, \bibinfo {author} {\bibfnamefont {T.}~\bibnamefont {Ngo}}, \bibinfo {author} {\bibfnamefont {E.}~\bibnamefont {Raimondo}}, \bibinfo {author} {\bibfnamefont {A.}~\bibnamefont {Giordano}}, \bibinfo {author} {\bibfnamefont {J.}~\bibnamefont {Igarashi}}, \bibinfo {author} {\bibfnamefont {B.}~\bibnamefont {Jinnai}}, \bibinfo {author} {\bibfnamefont {S.}~\bibnamefont {Zhao}}, \bibinfo {author} {\bibfnamefont {J.}~\bibnamefont {Lei}}, \bibinfo {author} {\bibfnamefont {Y.-X.}\ \bibnamefont {Guo}}, \bibinfo {author} {\bibfnamefont {G.}~\bibnamefont {Finocchio}}, \emph {et~al.},\ }\bibfield  {title} {\bibinfo {title} {Nanoscale spin rectifiers for harvesting ambient radiofrequency energy},\ }\href {https://doi.org/10.1038/s41928-024-01212-1} {\bibfield  {journal} {\bibinfo  {journal} {Nature Electronics}\ }\textbf {\bibinfo {volume} {7}},\ \bibinfo {pages} {567} (\bibinfo {year} {2024})}\BibitemShut {NoStop}%
\bibitem [{\citenamefont {Authors}(2020)}]{nature2020sttram}%
  \BibitemOpen
  \bibfield  {author} {\bibinfo {author} {\bibfnamefont {V.}~\bibnamefont {Authors}},\ }\bibfield  {title} {\bibinfo {title} {Neuromorphic spintronics},\ }\href {https://doi.org/10.1038/s41928-019-0360-9} {\bibfield  {journal} {\bibinfo  {journal} {Nature Electronics}\ }\textbf {\bibinfo {volume} {3}},\ \bibinfo {pages} {360} (\bibinfo {year} {2020})}\BibitemShut {NoStop}%
\bibitem [{\citenamefont {Dieny}\ \emph {et~al.}(2020)\citenamefont {Dieny}, \citenamefont {Prejbeanu}, \citenamefont {Garello}, \citenamefont {Gambardella}, \citenamefont {Freitas}, \citenamefont {Lehndorff}, \citenamefont {Raberg}, \citenamefont {Ebels}, \citenamefont {Demokritov}, \citenamefont {Akerman} \emph {et~al.}}]{dieny2020opportunities}%
  \BibitemOpen
  \bibfield  {author} {\bibinfo {author} {\bibfnamefont {B.}~\bibnamefont {Dieny}}, \bibinfo {author} {\bibfnamefont {I.~L.}\ \bibnamefont {Prejbeanu}}, \bibinfo {author} {\bibfnamefont {K.}~\bibnamefont {Garello}}, \bibinfo {author} {\bibfnamefont {P.}~\bibnamefont {Gambardella}}, \bibinfo {author} {\bibfnamefont {P.}~\bibnamefont {Freitas}}, \bibinfo {author} {\bibfnamefont {R.}~\bibnamefont {Lehndorff}}, \bibinfo {author} {\bibfnamefont {W.}~\bibnamefont {Raberg}}, \bibinfo {author} {\bibfnamefont {U.}~\bibnamefont {Ebels}}, \bibinfo {author} {\bibfnamefont {S.~O.}\ \bibnamefont {Demokritov}}, \bibinfo {author} {\bibfnamefont {J.}~\bibnamefont {Akerman}}, \emph {et~al.},\ }\bibfield  {title} {\bibinfo {title} {Opportunities and challenges for spintronics in the microelectronics industry},\ }\href {https://doi.org/https://doi.org/10.1038/s41928-020-0461-5} {\bibfield  {journal} {\bibinfo  {journal} {Nature Electronics}\ }\textbf {\bibinfo {volume} {3}},\ \bibinfo {pages} {446} (\bibinfo {year}
  {2020})}\BibitemShut {NoStop}%
\bibitem [{\citenamefont {Hu}\ \emph {et~al.}(2022)\citenamefont {Hu}, \citenamefont {Safranski},\ and\ \citenamefont {Sun}}]{hu2022dsmtj}%
  \BibitemOpen
  \bibfield  {author} {\bibinfo {author} {\bibfnamefont {G.}~\bibnamefont {Hu}}, \bibinfo {author} {\bibfnamefont {C.}~\bibnamefont {Safranski}},\ and\ \bibinfo {author} {\bibfnamefont {J.~Z.}\ \bibnamefont {Sun}},\ }\bibfield  {title} {\bibinfo {title} {Double spin-torque magnetic tunnel junction devices for last-level cache applications},\ }in\ \href {https://doi.org/10.1109/IEDM45625.2022.10019402} {\emph {\bibinfo {booktitle} {IEEE International Electron Devices Meeting}}}\ (\bibinfo {year} {2022})\BibitemShut {NoStop}%
\bibitem [{\citenamefont {Han}\ and\ \citenamefont {Jiang}(2024{\natexlab{a}})}]{han2024gpu}%
  \BibitemOpen
  \bibfield  {author} {\bibinfo {author} {\bibfnamefont {S.}~\bibnamefont {Han}}\ and\ \bibinfo {author} {\bibfnamefont {Y.}~\bibnamefont {Jiang}},\ }\bibfield  {title} {\bibinfo {title} {Advanced hybrid mram based novel gpu cache system for graphic processing with high efficiency},\ }\href {https://doi.org/10.1063/9.0000721} {\bibfield  {journal} {\bibinfo  {journal} {AIP Advances}\ }\textbf {\bibinfo {volume} {14}},\ \bibinfo {pages} {015058} (\bibinfo {year} {2024}{\natexlab{a}})}\BibitemShut {NoStop}%
\bibitem [{\citenamefont {Han}\ and\ \citenamefont {Jiang}(2024{\natexlab{b}})}]{han2024cpu}%
  \BibitemOpen
  \bibfield  {author} {\bibinfo {author} {\bibfnamefont {S.}~\bibnamefont {Han}}\ and\ \bibinfo {author} {\bibfnamefont {Y.}~\bibnamefont {Jiang}},\ }\bibfield  {title} {\bibinfo {title} {Novel cpu cache architecture based on two-dimensional mtj device with ferromagnetic fe3gete2},\ }\href {https://doi.org/10.1063/9.0000675} {\bibfield  {journal} {\bibinfo  {journal} {AIP Advances}\ }\textbf {\bibinfo {volume} {14}},\ \bibinfo {pages} {015027} (\bibinfo {year} {2024}{\natexlab{b}})}\BibitemShut {NoStop}%
\bibitem [{\citenamefont {Foundary}()}]{GF2023}%
  \BibitemOpen
  \bibfield  {author} {\bibinfo {author} {\bibfnamefont {G.}~\bibnamefont {Foundary}},\ }\href {https://gf.com/blog/making-new-memories-22nm-emram-ready-displace-eflash/} {\bibinfo {title} {{New Memories: 22nm eMRAM is Ready to Displace eFlash}}}\BibitemShut {NoStop}%
\bibitem [{\citenamefont {Shinoda}\ \emph {et~al.}(2024)\citenamefont {Shinoda}, \citenamefont {Igarashi}, \citenamefont {Jinnai}, \citenamefont {Fukami},\ and\ \citenamefont {Ohno}}]{10368042}%
  \BibitemOpen
  \bibfield  {author} {\bibinfo {author} {\bibfnamefont {T.}~\bibnamefont {Shinoda}}, \bibinfo {author} {\bibfnamefont {J.}~\bibnamefont {Igarashi}}, \bibinfo {author} {\bibfnamefont {B.}~\bibnamefont {Jinnai}}, \bibinfo {author} {\bibfnamefont {S.}~\bibnamefont {Fukami}},\ and\ \bibinfo {author} {\bibfnamefont {H.}~\bibnamefont {Ohno}},\ }\bibfield  {title} {\bibinfo {title} {Pitch scaling prospect of ultra-small magnetic tunnel junctions for high-density stt-mram: Effects of magnetostatic interference from neighboring bits},\ }\href {https://doi.org/10.1109/LED.2023.3345743} {\bibfield  {journal} {\bibinfo  {journal} {IEEE Electron Device Letters}\ }\textbf {\bibinfo {volume} {45}},\ \bibinfo {pages} {184} (\bibinfo {year} {2024})}\BibitemShut {NoStop}%
\bibitem [{\citenamefont {Watanabe}\ \emph {et~al.}(2018)\citenamefont {Watanabe}, \citenamefont {Jinnai}, \citenamefont {Fukami}, \citenamefont {Sato},\ and\ \citenamefont {Ohno}}]{watanabe2018shape}%
  \BibitemOpen
  \bibfield  {author} {\bibinfo {author} {\bibfnamefont {K.}~\bibnamefont {Watanabe}}, \bibinfo {author} {\bibfnamefont {B.}~\bibnamefont {Jinnai}}, \bibinfo {author} {\bibfnamefont {S.}~\bibnamefont {Fukami}}, \bibinfo {author} {\bibfnamefont {H.}~\bibnamefont {Sato}},\ and\ \bibinfo {author} {\bibfnamefont {H.}~\bibnamefont {Ohno}},\ }\bibfield  {title} {\bibinfo {title} {Shape anisotropy revisited in single-digit nanometer magnetic tunnel junctions},\ }\href@noop {} {\bibfield  {journal} {\bibinfo  {journal} {Nature communications}\ }\textbf {\bibinfo {volume} {9}},\ \bibinfo {pages} {663} (\bibinfo {year} {2018})}\BibitemShut {NoStop}%
\bibitem [{\citenamefont {Imec}(2022)}]{imec2022sotmram}%
  \BibitemOpen
  \bibfield  {author} {\bibinfo {author} {\bibnamefont {Imec}},\ }\href@noop {} {\bibinfo {title} {{SOT-MRAM for High-Density Last-Level Cache Memory}}},\ \bibinfo {howpublished} {\href{https://www.imec-int.com/en/articles/novel-sot-mram-architecture-opens-doors-high-density-last-level-cache-memory-applications}{Online article}} (\bibinfo {year} {2022}),\ \bibinfo {note} {accessed: 2024-10-30}\BibitemShut {NoStop}%
\bibitem [{\citenamefont {Nguyen}\ \emph {et~al.}(2024)\citenamefont {Nguyen}, \citenamefont {Rao}, \citenamefont {Wostyn},\ and\ \citenamefont {Couet}}]{nguyen2024sotmram}%
  \BibitemOpen
  \bibfield  {author} {\bibinfo {author} {\bibfnamefont {V.~D.}\ \bibnamefont {Nguyen}}, \bibinfo {author} {\bibfnamefont {S.}~\bibnamefont {Rao}}, \bibinfo {author} {\bibfnamefont {K.}~\bibnamefont {Wostyn}},\ and\ \bibinfo {author} {\bibfnamefont {S.}~\bibnamefont {Couet}},\ }\bibfield  {title} {\bibinfo {title} {Recent progress in spin-orbit torque magnetic random-access memory},\ }\href {https://doi.org/10.1038/s44306-024-00044-1} {\bibfield  {journal} {\bibinfo  {journal} {NPJ Spintronics}\ }\textbf {\bibinfo {volume} {2}},\ \bibinfo {pages} {48} (\bibinfo {year} {2024})}\BibitemShut {NoStop}%
\bibitem [{\citenamefont {Shao}(2024)}]{shao2024magnetic}%
  \BibitemOpen
  \bibfield  {author} {\bibinfo {author} {\bibfnamefont {Q.}~\bibnamefont {Shao}},\ }\href {https://doi.org/https://doi.org/10.1038/d41586-024-00576-w} {\bibinfo {title} {Magnetic whirlpools offer improved data storage}} (\bibinfo {year} {2024})\BibitemShut {NoStop}%
\bibitem [{\citenamefont {Rajib}\ \emph {et~al.}(2024)\citenamefont {Rajib}, \citenamefont {Bindal}, \citenamefont {Raj}, \citenamefont {Kaushik},\ and\ \citenamefont {Atulasimha}}]{rajib2024skyrmion}%
  \BibitemOpen
  \bibfield  {author} {\bibinfo {author} {\bibfnamefont {M.~M.}\ \bibnamefont {Rajib}}, \bibinfo {author} {\bibfnamefont {N.}~\bibnamefont {Bindal}}, \bibinfo {author} {\bibfnamefont {R.~K.}\ \bibnamefont {Raj}}, \bibinfo {author} {\bibfnamefont {B.~K.}\ \bibnamefont {Kaushik}},\ and\ \bibinfo {author} {\bibfnamefont {J.}~\bibnamefont {Atulasimha}},\ }\bibfield  {title} {\bibinfo {title} {Skyrmion-mediated nonvolatile ternary memory},\ }\href {https://doi.org/https://doi.org/10.1038/s41598-024-66853-w} {\bibfield  {journal} {\bibinfo  {journal} {Scientific Reports}\ }\textbf {\bibinfo {volume} {14}},\ \bibinfo {pages} {17199} (\bibinfo {year} {2024})}\BibitemShut {NoStop}%
\bibitem [{\citenamefont {Chen}\ \emph {et~al.}(2019)\citenamefont {Chen}, \citenamefont {Zhang}, \citenamefont {Deng}, \citenamefont {Yang}, \citenamefont {Lei}, \citenamefont {Zhang}, \citenamefont {Kang},\ and\ \citenamefont {Zhao}}]{8667872}%
  \BibitemOpen
  \bibfield  {author} {\bibinfo {author} {\bibfnamefont {X.}~\bibnamefont {Chen}}, \bibinfo {author} {\bibfnamefont {H.}~\bibnamefont {Zhang}}, \bibinfo {author} {\bibfnamefont {E.}~\bibnamefont {Deng}}, \bibinfo {author} {\bibfnamefont {M.}~\bibnamefont {Yang}}, \bibinfo {author} {\bibfnamefont {N.}~\bibnamefont {Lei}}, \bibinfo {author} {\bibfnamefont {Y.}~\bibnamefont {Zhang}}, \bibinfo {author} {\bibfnamefont {W.}~\bibnamefont {Kang}},\ and\ \bibinfo {author} {\bibfnamefont {W.}~\bibnamefont {Zhao}},\ }\bibfield  {title} {\bibinfo {title} {Sky-ram: Skyrmionic random access memory},\ }\href {https://doi.org/10.1109/LED.2019.2905425} {\bibfield  {journal} {\bibinfo  {journal} {IEEE Electron Device Letters}\ }\textbf {\bibinfo {volume} {40}},\ \bibinfo {pages} {722} (\bibinfo {year} {2019})}\BibitemShut {NoStop}%
\bibitem [{\citenamefont {Yuasa}\ \emph {et~al.}(2004)\citenamefont {Yuasa}, \citenamefont {Nagahama}, \citenamefont {Fukushima}, \citenamefont {Suzuki},\ and\ \citenamefont {Ando}}]{yuasa2004giant}%
  \BibitemOpen
  \bibfield  {author} {\bibinfo {author} {\bibfnamefont {S.}~\bibnamefont {Yuasa}}, \bibinfo {author} {\bibfnamefont {T.}~\bibnamefont {Nagahama}}, \bibinfo {author} {\bibfnamefont {A.}~\bibnamefont {Fukushima}}, \bibinfo {author} {\bibfnamefont {Y.}~\bibnamefont {Suzuki}},\ and\ \bibinfo {author} {\bibfnamefont {K.}~\bibnamefont {Ando}},\ }\bibfield  {title} {\bibinfo {title} {Giant room-temperature magnetoresistance in single-crystal fe/mgo/fe magnetic tunnel junctions},\ }\href {https://doi.org/https://doi.org/10.1038/nmat1257} {\bibfield  {journal} {\bibinfo  {journal} {Nature materials}\ }\textbf {\bibinfo {volume} {3}},\ \bibinfo {pages} {868} (\bibinfo {year} {2004})}\BibitemShut {NoStop}%
\bibitem [{\citenamefont {Scheike}\ \emph {et~al.}(2023{\natexlab{a}})\citenamefont {Scheike}, \citenamefont {Wen}, \citenamefont {Sukegawa},\ and\ \citenamefont {Mitani}}]{scheike2023room}%
  \BibitemOpen
  \bibfield  {author} {\bibinfo {author} {\bibfnamefont {T.}~\bibnamefont {Scheike}}, \bibinfo {author} {\bibfnamefont {Z.}~\bibnamefont {Wen}}, \bibinfo {author} {\bibfnamefont {H.}~\bibnamefont {Sukegawa}},\ and\ \bibinfo {author} {\bibfnamefont {S.}~\bibnamefont {Mitani}},\ }\bibfield  {title} {\bibinfo {title} {631\% room temperature tunnel magnetoresistance with large oscillation effect in cofe/mgo/cofe(001) junctions},\ }\href {https://doi.org/10.1063/5.0145873} {\bibfield  {journal} {\bibinfo  {journal} {Applied Physics Letters}\ }\textbf {\bibinfo {volume} {122}},\ \bibinfo {pages} {112404} (\bibinfo {year} {2023}{\natexlab{a}})}\BibitemShut {NoStop}%
\bibitem [{\citenamefont {Yang}\ \emph {et~al.}(2018)\citenamefont {Yang}, \citenamefont {Tao}, \citenamefont {Jiang}, \citenamefont {Chen}, \citenamefont {Tang}, \citenamefont {Yan},\ and\ \citenamefont {Han}}]{yang2018ultrahigh}%
  \BibitemOpen
  \bibfield  {author} {\bibinfo {author} {\bibfnamefont {B.}~\bibnamefont {Yang}}, \bibinfo {author} {\bibfnamefont {L.}~\bibnamefont {Tao}}, \bibinfo {author} {\bibfnamefont {L.}~\bibnamefont {Jiang}}, \bibinfo {author} {\bibfnamefont {W.}~\bibnamefont {Chen}}, \bibinfo {author} {\bibfnamefont {P.}~\bibnamefont {Tang}}, \bibinfo {author} {\bibfnamefont {Y.}~\bibnamefont {Yan}},\ and\ \bibinfo {author} {\bibfnamefont {X.}~\bibnamefont {Han}},\ }\bibfield  {title} {\bibinfo {title} {Ultrahigh tunneling-magnetoresistance ratios in nitride-based perpendicular magnetic tunnel junctions from first principles},\ }\href {https://doi.org/10.1103/PhysRevApplied.9.054019} {\bibfield  {journal} {\bibinfo  {journal} {Physical Review Applied}\ }\textbf {\bibinfo {volume} {9}},\ \bibinfo {pages} {054019} (\bibinfo {year} {2018})}\BibitemShut {NoStop}%
\bibitem [{\citenamefont {Balc{\i}}\ \emph {et~al.}(2018)\citenamefont {Balc{\i}}, \citenamefont {Akkus},\ and\ \citenamefont {Berber}}]{balci2018high}%
  \BibitemOpen
  \bibfield  {author} {\bibinfo {author} {\bibfnamefont {E.}~\bibnamefont {Balc{\i}}}, \bibinfo {author} {\bibfnamefont {U.~O.}\ \bibnamefont {Akkus}},\ and\ \bibinfo {author} {\bibfnamefont {S.}~\bibnamefont {Berber}},\ }\bibfield  {title} {\bibinfo {title} {High tmr in mxene-based mn2cf2/ti2co2/mn2cf2 magnetic tunneling junction},\ }\href@noop {} {\bibfield  {journal} {\bibinfo  {journal} {ACS Applied Materials \& Interfaces}\ }\textbf {\bibinfo {volume} {11}},\ \bibinfo {pages} {3609} (\bibinfo {year} {2018})}\BibitemShut {NoStop}%
\bibitem [{\citenamefont {Zhu}\ \emph {et~al.}(2023)\citenamefont {Zhu}, \citenamefont {Zhu}, \citenamefont {Zhou}, \citenamefont {Zhang}, \citenamefont {Lin}, \citenamefont {Cui}, \citenamefont {Yan}, \citenamefont {Wang}, \citenamefont {Deng}, \citenamefont {Yang} \emph {et~al.}}]{zhu2023large}%
  \BibitemOpen
  \bibfield  {author} {\bibinfo {author} {\bibfnamefont {W.}~\bibnamefont {Zhu}}, \bibinfo {author} {\bibfnamefont {Y.}~\bibnamefont {Zhu}}, \bibinfo {author} {\bibfnamefont {T.}~\bibnamefont {Zhou}}, \bibinfo {author} {\bibfnamefont {X.}~\bibnamefont {Zhang}}, \bibinfo {author} {\bibfnamefont {H.}~\bibnamefont {Lin}}, \bibinfo {author} {\bibfnamefont {Q.}~\bibnamefont {Cui}}, \bibinfo {author} {\bibfnamefont {F.}~\bibnamefont {Yan}}, \bibinfo {author} {\bibfnamefont {Z.}~\bibnamefont {Wang}}, \bibinfo {author} {\bibfnamefont {Y.}~\bibnamefont {Deng}}, \bibinfo {author} {\bibfnamefont {H.}~\bibnamefont {Yang}}, \emph {et~al.},\ }\bibfield  {title} {\bibinfo {title} {Large and tunable magnetoresistance in van der waals ferromagnet/semiconductor junctions},\ }\href {https://doi.org/10.1038/s41467-023-41077-0} {\bibfield  {journal} {\bibinfo  {journal} {Nature Communications}\ }\textbf {\bibinfo {volume} {14}},\ \bibinfo {pages} {1} (\bibinfo {year} {2023})}\BibitemShut {NoStop}%
\bibitem [{\citenamefont {Halder}\ \emph {et~al.}(2024)\citenamefont {Halder}, \citenamefont {Nell}, \citenamefont {Sihi}, \citenamefont {Bajaj}, \citenamefont {Sanvito},\ and\ \citenamefont {Droghetti}}]{halder2024half}%
  \BibitemOpen
  \bibfield  {author} {\bibinfo {author} {\bibfnamefont {A.}~\bibnamefont {Halder}}, \bibinfo {author} {\bibfnamefont {D.}~\bibnamefont {Nell}}, \bibinfo {author} {\bibfnamefont {A.}~\bibnamefont {Sihi}}, \bibinfo {author} {\bibfnamefont {A.}~\bibnamefont {Bajaj}}, \bibinfo {author} {\bibfnamefont {S.}~\bibnamefont {Sanvito}},\ and\ \bibinfo {author} {\bibfnamefont {A.}~\bibnamefont {Droghetti}},\ }\bibfield  {title} {\bibinfo {title} {Half-metallic transport and spin-polarized tunneling through the van der waals ferromagnet fe4gete2},\ }\href@noop {} {\bibfield  {journal} {\bibinfo  {journal} {Nano Letters}\ }\textbf {\bibinfo {volume} {24}},\ \bibinfo {pages} {9221} (\bibinfo {year} {2024})}\BibitemShut {NoStop}%
\bibitem [{\citenamefont {Nawa}\ \emph {et~al.}(2021)\citenamefont {Nawa}, \citenamefont {Masuda},\ and\ \citenamefont {Miura}}]{nawa2021enhanced}%
  \BibitemOpen
  \bibfield  {author} {\bibinfo {author} {\bibfnamefont {K.}~\bibnamefont {Nawa}}, \bibinfo {author} {\bibfnamefont {K.}~\bibnamefont {Masuda}},\ and\ \bibinfo {author} {\bibfnamefont {Y.}~\bibnamefont {Miura}},\ }\bibfield  {title} {\bibinfo {title} {Enhanced magnetoresistance under bias voltage in fe/mg o/mg al 2 o 4/mg o/fe trilayer tunneling barrier junction},\ }\href {https://doi.org/10.1103/PhysRevApplied.16.044037} {\bibfield  {journal} {\bibinfo  {journal} {Physical Review Applied}\ }\textbf {\bibinfo {volume} {16}},\ \bibinfo {pages} {044037} (\bibinfo {year} {2021})}\BibitemShut {NoStop}%
\bibitem [{\citenamefont {Lee}\ \emph {et~al.}(2022)\citenamefont {Lee}, \citenamefont {Lee}, \citenamefont {Kim}, \citenamefont {Oh}, \citenamefont {Lee}, \citenamefont {Jeong}, \citenamefont {Jang}, \citenamefont {Joo}, \citenamefont {Suh}, \citenamefont {Han} \emph {et~al.}}]{lee2022world}%
  \BibitemOpen
  \bibfield  {author} {\bibinfo {author} {\bibfnamefont {T.}~\bibnamefont {Lee}}, \bibinfo {author} {\bibfnamefont {J.}~\bibnamefont {Lee}}, \bibinfo {author} {\bibfnamefont {M.}~\bibnamefont {Kim}}, \bibinfo {author} {\bibfnamefont {J.}~\bibnamefont {Oh}}, \bibinfo {author} {\bibfnamefont {J.}~\bibnamefont {Lee}}, \bibinfo {author} {\bibfnamefont {H.}~\bibnamefont {Jeong}}, \bibinfo {author} {\bibfnamefont {P.}~\bibnamefont {Jang}}, \bibinfo {author} {\bibfnamefont {M.}~\bibnamefont {Joo}}, \bibinfo {author} {\bibfnamefont {K.}~\bibnamefont {Suh}}, \bibinfo {author} {\bibfnamefont {S.}~\bibnamefont {Han}}, \emph {et~al.},\ }\bibfield  {title} {\bibinfo {title} {World-most energy-efficient mram technology for non-volatile ram applications},\ }in\ \href@noop {} {\emph {\bibinfo {booktitle} {2022 International Electron Devices Meeting (IEDM)}}}\ (\bibinfo {organization} {IEEE},\ \bibinfo {year} {2022})\ pp.\ \bibinfo {pages} {10--7}\BibitemShut {NoStop}%
\bibitem [{\citenamefont {Fritsch}\ \emph {et~al.}(2006)\citenamefont {Fritsch}, \citenamefont {Schmidt},\ and\ \citenamefont {Grundmann}}]{10.1063/1.2188382}%
  \BibitemOpen
  \bibfield  {author} {\bibinfo {author} {\bibfnamefont {D.}~\bibnamefont {Fritsch}}, \bibinfo {author} {\bibfnamefont {H.}~\bibnamefont {Schmidt}},\ and\ \bibinfo {author} {\bibfnamefont {M.}~\bibnamefont {Grundmann}},\ }\bibfield  {title} {\bibinfo {title} {Pseudopotential band structures of rocksalt mgo, zno, and mg1-xznxo},\ }\href {https://doi.org/10.1063/1.2188382} {\bibfield  {journal} {\bibinfo  {journal} {Applied Physics Letters}\ }\textbf {\bibinfo {volume} {88}},\ \bibinfo {pages} {134104} (\bibinfo {year} {2006})}\BibitemShut {NoStop}%
\bibitem [{\citenamefont {Koster}\ \emph {et~al.}(2015)\citenamefont {Koster}, \citenamefont {Fang}, \citenamefont {Dijkstra}, \citenamefont {Van~Blaaderen},\ and\ \citenamefont {Van~Huis}}]{koster2015stabilization}%
  \BibitemOpen
  \bibfield  {author} {\bibinfo {author} {\bibfnamefont {R.~S.}\ \bibnamefont {Koster}}, \bibinfo {author} {\bibfnamefont {C.~M.}\ \bibnamefont {Fang}}, \bibinfo {author} {\bibfnamefont {M.}~\bibnamefont {Dijkstra}}, \bibinfo {author} {\bibfnamefont {A.}~\bibnamefont {Van~Blaaderen}},\ and\ \bibinfo {author} {\bibfnamefont {M.~A.}\ \bibnamefont {Van~Huis}},\ }\bibfield  {title} {\bibinfo {title} {Stabilization of rock salt zno nanocrystals by low-energy surfaces and mg additions: A first-principles study},\ }\href@noop {} {\bibfield  {journal} {\bibinfo  {journal} {The Journal of Physical Chemistry C}\ }\textbf {\bibinfo {volume} {119}},\ \bibinfo {pages} {5648} (\bibinfo {year} {2015})}\BibitemShut {NoStop}%
\bibitem [{\citenamefont {Lu}\ \emph {et~al.}(2016)\citenamefont {Lu}, \citenamefont {Tu}, \citenamefont {Yan}, \citenamefont {Trampert}, \citenamefont {Chang},\ and\ \citenamefont {Ploog}}]{lu2016growth}%
  \BibitemOpen
  \bibfield  {author} {\bibinfo {author} {\bibfnamefont {C.-Y.~J.}\ \bibnamefont {Lu}}, \bibinfo {author} {\bibfnamefont {Y.-T.}\ \bibnamefont {Tu}}, \bibinfo {author} {\bibfnamefont {T.}~\bibnamefont {Yan}}, \bibinfo {author} {\bibfnamefont {A.}~\bibnamefont {Trampert}}, \bibinfo {author} {\bibfnamefont {L.}~\bibnamefont {Chang}},\ and\ \bibinfo {author} {\bibfnamefont {K.}~\bibnamefont {Ploog}},\ }\bibfield  {title} {\bibinfo {title} {Growth and stability of rocksalt zn1-xmgxo epilayers and zno/mgo superlattice on mgo (100) substrate by molecular beam epitaxy},\ }\bibfield  {journal} {\bibinfo  {journal} {The Journal of Chemical Physics}\ }\textbf {\bibinfo {volume} {144}},\ \href {https://doi.org/https://doi.org/10.1063/1.4950885} {https://doi.org/10.1063/1.4950885} (\bibinfo {year} {2016})\BibitemShut {NoStop}%
\bibitem [{\citenamefont {Eijt}\ \emph {et~al.}(2007)\citenamefont {Eijt}, \citenamefont {De~Roode}, \citenamefont {Schut}, \citenamefont {Kooi},\ and\ \citenamefont {De~Hosson}}]{eijt2007formation}%
  \BibitemOpen
  \bibfield  {author} {\bibinfo {author} {\bibfnamefont {S.}~\bibnamefont {Eijt}}, \bibinfo {author} {\bibfnamefont {J.}~\bibnamefont {De~Roode}}, \bibinfo {author} {\bibfnamefont {H.}~\bibnamefont {Schut}}, \bibinfo {author} {\bibfnamefont {B.}~\bibnamefont {Kooi}},\ and\ \bibinfo {author} {\bibfnamefont {J.~T.~M.}\ \bibnamefont {De~Hosson}},\ }\bibfield  {title} {\bibinfo {title} {Formation and stability of rocksalt zno nanocrystals in mgo},\ }\bibfield  {journal} {\bibinfo  {journal} {Applied Physics Letters}\ }\textbf {\bibinfo {volume} {91}},\ \href {https://doi.org/10.1063/1.2809651} {10.1063/1.2809651} (\bibinfo {year} {2007})\BibitemShut {NoStop}%
\bibitem [{\citenamefont {Uehara}\ \emph {et~al.}(2010)\citenamefont {Uehara}, \citenamefont {Furuya}, \citenamefont {Sunaga}, \citenamefont {Miyajima},\ and\ \citenamefont {Kanai}}]{uehara2005mtj}%
  \BibitemOpen
  \bibfield  {author} {\bibinfo {author} {\bibfnamefont {Y.}~\bibnamefont {Uehara}}, \bibinfo {author} {\bibfnamefont {A.}~\bibnamefont {Furuya}}, \bibinfo {author} {\bibfnamefont {K.}~\bibnamefont {Sunaga}}, \bibinfo {author} {\bibfnamefont {T.}~\bibnamefont {Miyajima}},\ and\ \bibinfo {author} {\bibfnamefont {H.}~\bibnamefont {Kanai}},\ }\bibfield  {title} {\bibinfo {title} {Magnetic tunnel junctions with low resistance-area product of 0.5 $\omega\mu$m$^2$},\ }\href {https://www.jstage.jst.go.jp/article/msjmag/34/3/34_1003R027/_pdf/-char/ja} {\bibfield  {journal} {\bibinfo  {journal} {Journal of the Magnetic Society of Japan}\ } (\bibinfo {year} {2010})},\ \bibinfo {note} {next Generation Technical Computing Unit, Simulation Technology Development Division, Devices and Materials Laboratories}\BibitemShut {NoStop}%
\bibitem [{\citenamefont {Saito}\ \emph {et~al.}(2019)\citenamefont {Saito}, \citenamefont {Narayananellore}, \citenamefont {Matsuo}, \citenamefont {Doko}, \citenamefont {Kon}, \citenamefont {Yasukawa}, \citenamefont {Imamura},\ and\ \citenamefont {Yuasa}}]{PhysRevApplied.11.064032}%
  \BibitemOpen
  \bibfield  {author} {\bibinfo {author} {\bibfnamefont {H.}~\bibnamefont {Saito}}, \bibinfo {author} {\bibfnamefont {S.~K.}\ \bibnamefont {Narayananellore}}, \bibinfo {author} {\bibfnamefont {N.}~\bibnamefont {Matsuo}}, \bibinfo {author} {\bibfnamefont {N.}~\bibnamefont {Doko}}, \bibinfo {author} {\bibfnamefont {S.}~\bibnamefont {Kon}}, \bibinfo {author} {\bibfnamefont {Y.}~\bibnamefont {Yasukawa}}, \bibinfo {author} {\bibfnamefont {H.}~\bibnamefont {Imamura}},\ and\ \bibinfo {author} {\bibfnamefont {S.}~\bibnamefont {Yuasa}},\ }\bibfield  {title} {\bibinfo {title} {Tunneling magnetoresistance and spin-dependent diode performance in fully epitaxial magnetic tunnel junctions with a rocksalt $\mathrm{Zn}\mathrm{O}/\mathrm{Mg}\mathrm{O}$ bilayer tunnel barrier},\ }\href {https://doi.org/10.1103/PhysRevApplied.11.064032} {\bibfield  {journal} {\bibinfo  {journal} {Phys. Rev. Appl.}\ }\textbf {\bibinfo {volume} {11}},\ \bibinfo {pages} {064032} (\bibinfo {year} {2019})}\BibitemShut {NoStop}%
\bibitem [{\citenamefont {Soler}\ \emph {et~al.}(2002)\citenamefont {Soler}, \citenamefont {Artacho}, \citenamefont {Gale}, \citenamefont {Garc{\'i}a}, \citenamefont {Junquera}, \citenamefont {Ordej{\'o}n},\ and\ \citenamefont {S{\'a}nchez-Portal}}]{Soler2002}%
  \BibitemOpen
  \bibfield  {author} {\bibinfo {author} {\bibfnamefont {J.~M.}\ \bibnamefont {Soler}}, \bibinfo {author} {\bibfnamefont {E.}~\bibnamefont {Artacho}}, \bibinfo {author} {\bibfnamefont {J.~D.}\ \bibnamefont {Gale}}, \bibinfo {author} {\bibfnamefont {A.}~\bibnamefont {Garc{\'i}a}}, \bibinfo {author} {\bibfnamefont {J.}~\bibnamefont {Junquera}}, \bibinfo {author} {\bibfnamefont {P.}~\bibnamefont {Ordej{\'o}n}},\ and\ \bibinfo {author} {\bibfnamefont {D.}~\bibnamefont {S{\'a}nchez-Portal}},\ }\bibfield  {title} {\bibinfo {title} {The siesta method for \emph{ab initio} order-{N} materials simulation},\ }\href {https://doi.org/10.1088/0953-8984/14/11/302} {\bibfield  {journal} {\bibinfo  {journal} {Journal of Physics: Condensed Matter}\ }\textbf {\bibinfo {volume} {14}},\ \bibinfo {pages} {2745} (\bibinfo {year} {2002})}\BibitemShut {NoStop}%
\bibitem [{\citenamefont {{van Setten}}\ \emph {et~al.}(2018)\citenamefont {{van Setten}}, \citenamefont {Giantomassi}, \citenamefont {Bousquet}, \citenamefont {Verstraete}, \citenamefont {Hamann}, \citenamefont {Gonze},\ and\ \citenamefont {Rignanese}}]{VANSETTEN201839}%
  \BibitemOpen
  \bibfield  {author} {\bibinfo {author} {\bibfnamefont {M.}~\bibnamefont {{van Setten}}}, \bibinfo {author} {\bibfnamefont {M.}~\bibnamefont {Giantomassi}}, \bibinfo {author} {\bibfnamefont {E.}~\bibnamefont {Bousquet}}, \bibinfo {author} {\bibfnamefont {M.}~\bibnamefont {Verstraete}}, \bibinfo {author} {\bibfnamefont {D.}~\bibnamefont {Hamann}}, \bibinfo {author} {\bibfnamefont {X.}~\bibnamefont {Gonze}},\ and\ \bibinfo {author} {\bibfnamefont {G.-M.}\ \bibnamefont {Rignanese}},\ }\bibfield  {title} {\bibinfo {title} {The pseudodojo: Training and grading a 85 element optimized norm-conserving pseudopotential table},\ }\href {https://doi.org/https://doi.org/10.1016/j.cpc.2018.01.012} {\bibfield  {journal} {\bibinfo  {journal} {Computer Physics Communications}\ }\textbf {\bibinfo {volume} {226}},\ \bibinfo {pages} {39} (\bibinfo {year} {2018})}\BibitemShut {NoStop}%
\bibitem [{\citenamefont {Papior}(2024)}]{zerothi_sisl}%
  \BibitemOpen
  \bibfield  {author} {\bibinfo {author} {\bibfnamefont {N.}~\bibnamefont {Papior}},\ }\href {https://doi.org/10.5281/zenodo.597181} {\bibinfo {title} {sisl: v 0.16.0}} (\bibinfo {year} {2024})\BibitemShut {NoStop}%
\bibitem [{\citenamefont {Paulsson}\ and\ \citenamefont {Brandbyge}(2007)}]{Madseigen}%
  \BibitemOpen
  \bibfield  {author} {\bibinfo {author} {\bibfnamefont {M.}~\bibnamefont {Paulsson}}\ and\ \bibinfo {author} {\bibfnamefont {M.}~\bibnamefont {Brandbyge}},\ }\bibfield  {title} {\bibinfo {title} {Transmission eigenchannels from nonequilibrium green's functions},\ }\href {https://doi.org/10.1103/PhysRevB.76.115117} {\bibfield  {journal} {\bibinfo  {journal} {Phys. Rev. B}\ }\textbf {\bibinfo {volume} {76}},\ \bibinfo {pages} {115117} (\bibinfo {year} {2007})}\BibitemShut {NoStop}%
\bibitem [{\citenamefont {Baez~Flores}\ \emph {et~al.}(2024)\citenamefont {Baez~Flores}, \citenamefont {van Schilfgaarde},\ and\ \citenamefont {Belashchenko}}]{PhysRevB.110.224425}%
  \BibitemOpen
  \bibfield  {author} {\bibinfo {author} {\bibfnamefont {G.~G.}\ \bibnamefont {Baez~Flores}}, \bibinfo {author} {\bibfnamefont {M.}~\bibnamefont {van Schilfgaarde}},\ and\ \bibinfo {author} {\bibfnamefont {K.~D.}\ \bibnamefont {Belashchenko}},\ }\bibfield  {title} {\bibinfo {title} {Tunneling magnetoresistance in mgo tunnel junctions with fe-based leads in empirically corrected density functional theory},\ }\href {https://doi.org/10.1103/PhysRevB.110.224425} {\bibfield  {journal} {\bibinfo  {journal} {Phys. Rev. B}\ }\textbf {\bibinfo {volume} {110}},\ \bibinfo {pages} {224425} (\bibinfo {year} {2024})}\BibitemShut {NoStop}%
\bibitem [{\citenamefont {Faleev}\ \emph {et~al.}(2012)\citenamefont {Faleev}, \citenamefont {Mryasov},\ and\ \citenamefont {van Schilfgaarde}}]{PhysRevB.85.174433}%
  \BibitemOpen
  \bibfield  {author} {\bibinfo {author} {\bibfnamefont {S.~V.}\ \bibnamefont {Faleev}}, \bibinfo {author} {\bibfnamefont {O.~N.}\ \bibnamefont {Mryasov}},\ and\ \bibinfo {author} {\bibfnamefont {M.}~\bibnamefont {van Schilfgaarde}},\ }\bibfield  {title} {\bibinfo {title} {Effect of correlations on electronic structure and transport across (001) fe/mgo/fe junctions},\ }\href {https://doi.org/10.1103/PhysRevB.85.174433} {\bibfield  {journal} {\bibinfo  {journal} {Phys. Rev. B}\ }\textbf {\bibinfo {volume} {85}},\ \bibinfo {pages} {174433} (\bibinfo {year} {2012})}\BibitemShut {NoStop}%
\bibitem [{\citenamefont {Niaouris}\ \emph {et~al.}(2022)\citenamefont {Niaouris}, \citenamefont {Durnev}, \citenamefont {Linpeng}, \citenamefont {Viitaniemi}, \citenamefont {Zimmermann}, \citenamefont {Vishnuradhan}, \citenamefont {Kozuka}, \citenamefont {Kawasaki},\ and\ \citenamefont {Fu}}]{PhysRevB.105.195202}%
  \BibitemOpen
  \bibfield  {author} {\bibinfo {author} {\bibfnamefont {V.}~\bibnamefont {Niaouris}}, \bibinfo {author} {\bibfnamefont {M.~V.}\ \bibnamefont {Durnev}}, \bibinfo {author} {\bibfnamefont {X.}~\bibnamefont {Linpeng}}, \bibinfo {author} {\bibfnamefont {M.~L.~K.}\ \bibnamefont {Viitaniemi}}, \bibinfo {author} {\bibfnamefont {C.}~\bibnamefont {Zimmermann}}, \bibinfo {author} {\bibfnamefont {A.}~\bibnamefont {Vishnuradhan}}, \bibinfo {author} {\bibfnamefont {Y.}~\bibnamefont {Kozuka}}, \bibinfo {author} {\bibfnamefont {M.}~\bibnamefont {Kawasaki}},\ and\ \bibinfo {author} {\bibfnamefont {K.-M.~C.}\ \bibnamefont {Fu}},\ }\bibfield  {title} {\bibinfo {title} {Ensemble spin relaxation of shallow donor qubits in zno},\ }\href {https://doi.org/10.1103/PhysRevB.105.195202} {\bibfield  {journal} {\bibinfo  {journal} {Phys. Rev. B}\ }\textbf {\bibinfo {volume} {105}},\ \bibinfo {pages} {195202} (\bibinfo {year} {2022})}\BibitemShut {NoStop}%
\bibitem [{\citenamefont {Miwa}\ \emph {et~al.}(2017)\citenamefont {Miwa}, \citenamefont {Fujimoto}, \citenamefont {Risius}, \citenamefont {Nawaoka}, \citenamefont {Goto},\ and\ \citenamefont {Suzuki}}]{PhysRevX.7.031018}%
  \BibitemOpen
  \bibfield  {author} {\bibinfo {author} {\bibfnamefont {S.}~\bibnamefont {Miwa}}, \bibinfo {author} {\bibfnamefont {J.}~\bibnamefont {Fujimoto}}, \bibinfo {author} {\bibfnamefont {P.}~\bibnamefont {Risius}}, \bibinfo {author} {\bibfnamefont {K.}~\bibnamefont {Nawaoka}}, \bibinfo {author} {\bibfnamefont {M.}~\bibnamefont {Goto}},\ and\ \bibinfo {author} {\bibfnamefont {Y.}~\bibnamefont {Suzuki}},\ }\bibfield  {title} {\bibinfo {title} {Strong bias effect on voltage-driven torque at epitaxial fe-mgo interface},\ }\href {https://doi.org/10.1103/PhysRevX.7.031018} {\bibfield  {journal} {\bibinfo  {journal} {Phys. Rev. X}\ }\textbf {\bibinfo {volume} {7}},\ \bibinfo {pages} {031018} (\bibinfo {year} {2017})}\BibitemShut {NoStop}%
\bibitem [{\citenamefont {Huang}\ \emph {et~al.}(2021)\citenamefont {Huang}, \citenamefont {Chao},\ and\ \citenamefont {Tang}}]{huang2021thickness}%
  \BibitemOpen
  \bibfield  {author} {\bibinfo {author} {\bibfnamefont {B.-H.}\ \bibnamefont {Huang}}, \bibinfo {author} {\bibfnamefont {C.-C.}\ \bibnamefont {Chao}},\ and\ \bibinfo {author} {\bibfnamefont {Y.-H.}\ \bibnamefont {Tang}},\ }\bibfield  {title} {\bibinfo {title} {Thickness dependence of spin torque effect in fe/mgo/fe magnetic tunnel junction: Implementation of divide-and-conquer with first-principles calculation},\ }\href@noop {} {\bibfield  {journal} {\bibinfo  {journal} {AIP Advances}\ }\textbf {\bibinfo {volume} {11}} (\bibinfo {year} {2021})}\BibitemShut {NoStop}%
\bibitem [{\citenamefont {Brandbyge}\ \emph {et~al.}(2002)\citenamefont {Brandbyge}, \citenamefont {Mozos}, \citenamefont {Ordej\'on}, \citenamefont {Taylor},\ and\ \citenamefont {Stokbro}}]{PhysRevB.65.165401}%
  \BibitemOpen
  \bibfield  {author} {\bibinfo {author} {\bibfnamefont {M.}~\bibnamefont {Brandbyge}}, \bibinfo {author} {\bibfnamefont {J.-L.}\ \bibnamefont {Mozos}}, \bibinfo {author} {\bibfnamefont {P.}~\bibnamefont {Ordej\'on}}, \bibinfo {author} {\bibfnamefont {J.}~\bibnamefont {Taylor}},\ and\ \bibinfo {author} {\bibfnamefont {K.}~\bibnamefont {Stokbro}},\ }\bibfield  {title} {\bibinfo {title} {Density-functional method for nonequilibrium electron transport},\ }\href {https://doi.org/10.1103/PhysRevB.65.165401} {\bibfield  {journal} {\bibinfo  {journal} {Phys. Rev. B}\ }\textbf {\bibinfo {volume} {65}},\ \bibinfo {pages} {165401} (\bibinfo {year} {2002})}\BibitemShut {NoStop}%
\bibitem [{\citenamefont {Deac}\ \emph {et~al.}(2008)\citenamefont {Deac}, \citenamefont {Fukushima}, \citenamefont {Kubota}, \citenamefont {Maehara}, \citenamefont {Suzuki}, \citenamefont {Yuasa}, \citenamefont {Nagamine}, \citenamefont {Tsunekawa}, \citenamefont {Djayaprawira},\ and\ \citenamefont {Watanabe}}]{deac2008bias}%
  \BibitemOpen
  \bibfield  {author} {\bibinfo {author} {\bibfnamefont {A.~M.}\ \bibnamefont {Deac}}, \bibinfo {author} {\bibfnamefont {A.}~\bibnamefont {Fukushima}}, \bibinfo {author} {\bibfnamefont {H.}~\bibnamefont {Kubota}}, \bibinfo {author} {\bibfnamefont {H.}~\bibnamefont {Maehara}}, \bibinfo {author} {\bibfnamefont {Y.}~\bibnamefont {Suzuki}}, \bibinfo {author} {\bibfnamefont {S.}~\bibnamefont {Yuasa}}, \bibinfo {author} {\bibfnamefont {Y.}~\bibnamefont {Nagamine}}, \bibinfo {author} {\bibfnamefont {K.}~\bibnamefont {Tsunekawa}}, \bibinfo {author} {\bibfnamefont {D.~D.}\ \bibnamefont {Djayaprawira}},\ and\ \bibinfo {author} {\bibfnamefont {N.}~\bibnamefont {Watanabe}},\ }\bibfield  {title} {\bibinfo {title} {Bias-driven high-power microwave emission from mgo-based tunnel magnetoresistance devices},\ }\href {https://doi.org/10.1038/nphys1036} {\bibfield  {journal} {\bibinfo  {journal} {Nature Physics}\ }\textbf {\bibinfo {volume} {4}},\ \bibinfo {pages} {803} (\bibinfo {year} {2008})}\BibitemShut {NoStop}%
\bibitem [{\citenamefont {Patel}\ \emph {et~al.}(2020)\citenamefont {Patel}, \citenamefont {Jan},\ and\ \citenamefont {Wang}}]{patel2020mgo}%
  \BibitemOpen
  \bibfield  {author} {\bibinfo {author} {\bibfnamefont {S.}~\bibnamefont {Patel}}, \bibinfo {author} {\bibfnamefont {G.}~\bibnamefont {Jan}},\ and\ \bibinfo {author} {\bibfnamefont {Y.-J.}\ \bibnamefont {Wang}},\ }\href {https://patents.google.com/patent/US20200270737A1/en} {\bibinfo {title} {Monolayer-by-monolayer growth of mgo layers using mg sublimation and oxidation}} (\bibinfo {year} {2020}),\ \bibinfo {note} {u.S. Patent Application Publication}\BibitemShut {NoStop}%
\bibitem [{\citenamefont {Wentzcovitch}\ \emph {et~al.}(1992)\citenamefont {Wentzcovitch}, \citenamefont {Martins},\ and\ \citenamefont {Allen}}]{wentzcovitch1992energy}%
  \BibitemOpen
  \bibfield  {author} {\bibinfo {author} {\bibfnamefont {R.~M.}\ \bibnamefont {Wentzcovitch}}, \bibinfo {author} {\bibfnamefont {J.~L.}\ \bibnamefont {Martins}},\ and\ \bibinfo {author} {\bibfnamefont {P.~B.}\ \bibnamefont {Allen}},\ }\bibfield  {title} {\bibinfo {title} {Energy versus free-energy conservation in first-principles molecular dynamics},\ }\href {https://doi.org/10.1103/PhysRevB.45.11372} {\bibfield  {journal} {\bibinfo  {journal} {Physical Review B}\ }\textbf {\bibinfo {volume} {45}},\ \bibinfo {pages} {11372} (\bibinfo {year} {1992})}\BibitemShut {NoStop}%
\bibitem [{\citenamefont {Matsumoto}\ \emph {et~al.}(2007)\citenamefont {Matsumoto}, \citenamefont {Fukushima}, \citenamefont {Nagahama}, \citenamefont {Suzuki}, \citenamefont {Ando},\ and\ \citenamefont {Yuasa}}]{matsumoto2007oscillation}%
  \BibitemOpen
  \bibfield  {author} {\bibinfo {author} {\bibfnamefont {R.}~\bibnamefont {Matsumoto}}, \bibinfo {author} {\bibfnamefont {A.}~\bibnamefont {Fukushima}}, \bibinfo {author} {\bibfnamefont {T.}~\bibnamefont {Nagahama}}, \bibinfo {author} {\bibfnamefont {Y.}~\bibnamefont {Suzuki}}, \bibinfo {author} {\bibfnamefont {K.}~\bibnamefont {Ando}},\ and\ \bibinfo {author} {\bibfnamefont {S.}~\bibnamefont {Yuasa}},\ }\bibfield  {title} {\bibinfo {title} {Oscillation of giant tunneling magnetoresistance with respect to tunneling barrier thickness in fully epitaxial fe/ mgo/ fe magnetic tunnel junctions},\ }\href@noop {} {\bibfield  {journal} {\bibinfo  {journal} {Applied physics letters}\ }\textbf {\bibinfo {volume} {90}} (\bibinfo {year} {2007})}\BibitemShut {NoStop}%
\bibitem [{\citenamefont {Scheike}\ \emph {et~al.}(2023{\natexlab{b}})\citenamefont {Scheike}, \citenamefont {Wen}, \citenamefont {Sukegawa},\ and\ \citenamefont {Mitani}}]{scheike2023631}%
  \BibitemOpen
  \bibfield  {author} {\bibinfo {author} {\bibfnamefont {T.}~\bibnamefont {Scheike}}, \bibinfo {author} {\bibfnamefont {Z.}~\bibnamefont {Wen}}, \bibinfo {author} {\bibfnamefont {H.}~\bibnamefont {Sukegawa}},\ and\ \bibinfo {author} {\bibfnamefont {S.}~\bibnamefont {Mitani}},\ }\bibfield  {title} {\bibinfo {title} {631\% room temperature tunnel magnetoresistance with large oscillation effect in cofe/mgo/cofe (001) junctions},\ }\href@noop {} {\bibfield  {journal} {\bibinfo  {journal} {Applied Physics Letters}\ }\textbf {\bibinfo {volume} {122}} (\bibinfo {year} {2023}{\natexlab{b}})}\BibitemShut {NoStop}%
\bibitem [{\citenamefont {Mermin}(1965)}]{mermin1965thermal}%
  \BibitemOpen
  \bibfield  {author} {\bibinfo {author} {\bibfnamefont {N.~D.}\ \bibnamefont {Mermin}},\ }\bibfield  {title} {\bibinfo {title} {Thermal properties of the inhomogeneous electron gas},\ }\href {https://doi.org/10.1103/PhysRev.137.A1441} {\bibfield  {journal} {\bibinfo  {journal} {Physical Review}\ }\textbf {\bibinfo {volume} {137}},\ \bibinfo {pages} {A1441} (\bibinfo {year} {1965})}\BibitemShut {NoStop}%
\bibitem [{\citenamefont {Jiang}\ \emph {et~al.}(2010)\citenamefont {Jiang}, \citenamefont {Gomez-Abal}, \citenamefont {Rinke},\ and\ \citenamefont {Scheffler}}]{jiang2010first}%
  \BibitemOpen
  \bibfield  {author} {\bibinfo {author} {\bibfnamefont {H.}~\bibnamefont {Jiang}}, \bibinfo {author} {\bibfnamefont {R.~I.}\ \bibnamefont {Gomez-Abal}}, \bibinfo {author} {\bibfnamefont {P.}~\bibnamefont {Rinke}},\ and\ \bibinfo {author} {\bibfnamefont {M.}~\bibnamefont {Scheffler}},\ }\bibfield  {title} {\bibinfo {title} {First-principles modeling of localized d states with the gw@ lda+ u approach},\ }\href@noop {} {\bibfield  {journal} {\bibinfo  {journal} {Physical Review B—Condensed Matter and Materials Physics}\ }\textbf {\bibinfo {volume} {82}},\ \bibinfo {pages} {045108} (\bibinfo {year} {2010})}\BibitemShut {NoStop}%
\bibitem [{\citenamefont {Sch{\"o}nberger}\ and\ \citenamefont {Aryasetiawan}(1995)}]{schonberger1995bulk}%
  \BibitemOpen
  \bibfield  {author} {\bibinfo {author} {\bibfnamefont {U.}~\bibnamefont {Sch{\"o}nberger}}\ and\ \bibinfo {author} {\bibfnamefont {F.}~\bibnamefont {Aryasetiawan}},\ }\bibfield  {title} {\bibinfo {title} {Bulk and surface electronic structures of mgo},\ }\href@noop {} {\bibfield  {journal} {\bibinfo  {journal} {Physical Review B}\ }\textbf {\bibinfo {volume} {52}},\ \bibinfo {pages} {8788} (\bibinfo {year} {1995})}\BibitemShut {NoStop}%
\bibitem [{\citenamefont {Haule}\ and\ \citenamefont {Mandal}(2024)}]{haule2024all}%
  \BibitemOpen
  \bibfield  {author} {\bibinfo {author} {\bibfnamefont {K.}~\bibnamefont {Haule}}\ and\ \bibinfo {author} {\bibfnamefont {S.}~\bibnamefont {Mandal}},\ }\bibfield  {title} {\bibinfo {title} {All electron gw with linearized augmented plane waves for metals and semiconductors},\ }\href@noop {} {\bibfield  {journal} {\bibinfo  {journal} {Computer Physics Communications}\ }\textbf {\bibinfo {volume} {295}},\ \bibinfo {pages} {108986} (\bibinfo {year} {2024})}\BibitemShut {NoStop}%
\bibitem [{\citenamefont {Zermatten}\ \emph {et~al.}(2008)\citenamefont {Zermatten}, \citenamefont {Gaudin}, \citenamefont {Maris}, \citenamefont {Miron}, \citenamefont {Schuhl}, \citenamefont {Tiusan}, \citenamefont {Greullet},\ and\ \citenamefont {Hehn}}]{PhysRevB.78.033301}%
  \BibitemOpen
  \bibfield  {author} {\bibinfo {author} {\bibfnamefont {P.-J.}\ \bibnamefont {Zermatten}}, \bibinfo {author} {\bibfnamefont {G.}~\bibnamefont {Gaudin}}, \bibinfo {author} {\bibfnamefont {G.}~\bibnamefont {Maris}}, \bibinfo {author} {\bibfnamefont {M.}~\bibnamefont {Miron}}, \bibinfo {author} {\bibfnamefont {A.}~\bibnamefont {Schuhl}}, \bibinfo {author} {\bibfnamefont {C.}~\bibnamefont {Tiusan}}, \bibinfo {author} {\bibfnamefont {F.}~\bibnamefont {Greullet}},\ and\ \bibinfo {author} {\bibfnamefont {M.}~\bibnamefont {Hehn}},\ }\bibfield  {title} {\bibinfo {title} {Experimental evidence of interface resonance states in single-crystal magnetic tunnel junctions},\ }\href {https://doi.org/10.1103/PhysRevB.78.033301} {\bibfield  {journal} {\bibinfo  {journal} {Phys. Rev. B}\ }\textbf {\bibinfo {volume} {78}},\ \bibinfo {pages} {033301} (\bibinfo {year} {2008})}\BibitemShut {NoStop}%
\bibitem [{\citenamefont {Du}\ \emph {et~al.}(2010)\citenamefont {Du}, \citenamefont {Wang}, \citenamefont {Ma}, \citenamefont {Wang}, \citenamefont {Ward}, \citenamefont {Zhang}, \citenamefont {Wang}, \citenamefont {Kohn},\ and\ \citenamefont {Han}}]{PhysRevB.81.064438}%
  \BibitemOpen
  \bibfield  {author} {\bibinfo {author} {\bibfnamefont {G.~X.}\ \bibnamefont {Du}}, \bibinfo {author} {\bibfnamefont {S.~G.}\ \bibnamefont {Wang}}, \bibinfo {author} {\bibfnamefont {Q.~L.}\ \bibnamefont {Ma}}, \bibinfo {author} {\bibfnamefont {Y.}~\bibnamefont {Wang}}, \bibinfo {author} {\bibfnamefont {R.~C.~C.}\ \bibnamefont {Ward}}, \bibinfo {author} {\bibfnamefont {X.-G.}\ \bibnamefont {Zhang}}, \bibinfo {author} {\bibfnamefont {C.}~\bibnamefont {Wang}}, \bibinfo {author} {\bibfnamefont {A.}~\bibnamefont {Kohn}},\ and\ \bibinfo {author} {\bibfnamefont {X.~F.}\ \bibnamefont {Han}},\ }\bibfield  {title} {\bibinfo {title} {Spin-dependent tunneling spectroscopy for interface characterization of epitaxial fe/mgo/fe magnetic tunnel junctions},\ }\href {https://doi.org/10.1103/PhysRevB.81.064438} {\bibfield  {journal} {\bibinfo  {journal} {Phys. Rev. B}\ }\textbf {\bibinfo {volume} {81}},\ \bibinfo {pages} {064438} (\bibinfo {year} {2010})}\BibitemShut {NoStop}%
\bibitem [{\citenamefont {B\"uttiker}\ \emph {et~al.}(1985)\citenamefont {B\"uttiker}, \citenamefont {Imry}, \citenamefont {Landauer},\ and\ \citenamefont {Pinhas}}]{PhysRevB.31.6207}%
  \BibitemOpen
  \bibfield  {author} {\bibinfo {author} {\bibfnamefont {M.}~\bibnamefont {B\"uttiker}}, \bibinfo {author} {\bibfnamefont {Y.}~\bibnamefont {Imry}}, \bibinfo {author} {\bibfnamefont {R.}~\bibnamefont {Landauer}},\ and\ \bibinfo {author} {\bibfnamefont {S.}~\bibnamefont {Pinhas}},\ }\bibfield  {title} {\bibinfo {title} {Generalized many-channel conductance formula with application to small rings},\ }\href {https://doi.org/10.1103/PhysRevB.31.6207} {\bibfield  {journal} {\bibinfo  {journal} {Phys. Rev. B}\ }\textbf {\bibinfo {volume} {31}},\ \bibinfo {pages} {6207} (\bibinfo {year} {1985})}\BibitemShut {NoStop}%
\bibitem [{\citenamefont {Datta}(2005)}]{datta2005quantum}%
  \BibitemOpen
  \bibfield  {author} {\bibinfo {author} {\bibfnamefont {S.}~\bibnamefont {Datta}},\ }\href@noop {} {\emph {\bibinfo {title} {Quantum transport: atom to transistor}}}\ (\bibinfo  {publisher} {Cambridge university press},\ \bibinfo {year} {2005})\BibitemShut {NoStop}%
\end{thebibliography}%

\end{document}